\documentstyle[aps,epsf,prb,eqsecnum]{revtex}

\begin{document}
\title{Spontaneous Hall effect in chiral $p$-wave superconductor}

\author{Akira Furusaki,$^1$ Masashige Matsumoto,$^2$
 and Manfred Sigrist$^1$}
\address{
$^1$Yukawa Institute for Theoretical Physics, Kyoto University, Kyoto
 606-8502, Japan\\
$^2$Department of Physics, Shizuoka University, Shizuoka 422-8529,
 Japan} 

\date{February 8, 2001}
\maketitle

\begin{abstract}
In a chiral superconductor with broken time-reversal symmetry
a ``spontaneous Hall effect'' may be observed.
We analyze this phenomenon by taking into account the surface
properties of a chiral superconductor.
We identify two main contributions to the spontaneous Hall effect.
One contribution originates from the Bernoulli (or Lorentz) force
from spontaneous currents running along the surfaces of the
superconductor.
The other contribution has a topological origin and is related to the
intrinsic angular momentum of Cooper pairs.
The latter can be described in terms of a Chern-Simons-like term in
the low-energy field theory of the superconductor and has some
similarities with the quantum Hall effect.
The spontaneous Hall effect in a chiral superconductor
is, however, non-universal.
Our analysis is based on three approaches to the problem:
a self-consistent solution of the Bogoliubov-de Gennes equation,
a generalized Ginzburg-Landau theory, and a hydrodynamic formulation.
All three methods consistently lead to the same conclusion that the
spontaneous Hall resistance of a two-dimensional superconducting Hall
bar is of order $h/(e k_F\lambda)^2$, where $k_F$ is the Fermi wave
vector and $\lambda$ is the London penetration depth;
the Hall resistance is substantially suppressed from a quantum unit of
resistance.
Experimental issues in measuring this effect are briefly discussed. 
\end{abstract}



\section{Introduction}
Unconventional superconductivity appears with a large variety of
possible phases displaying many properties that are not shared by
conventional $s$-wave superconductors.
These phases are characterized by their symmetry properties, and in
most cases not only the U(1)-gauge symmetry but other symmetries
are also spontaneously broken.
Interesting physics emerges, in particular, when
time-reversal symmetry ${\cal T}$ is violated in a superconductor.
Volovik and Gor'kov have classified such superconducting states into
two categories, the so-called ``ferromagnetic'' and the
``antiferromagnetic'' states.\cite{VG}
They are distinguished by the internal angular momentum of Cooper
pairs.
In the ferromagnetic state the Cooper pairs possess either a finite
orbital or (for non-unitary states) spin moment, while in the
antiferromagnetic state they have no net moments.
Examples of these two types of states were recently
discussed in connection with high-temperature superconductors;
the so-called $d_{x^2-y^2} + i d_{xy}$-wave state represents a
ferromagnetic pairing state, while the $ d_{x^2-y^2} + i s $-wave
state is antiferromagnetic. In high-temperature superconductors a $
{\cal T} $-violating state is most likely to be realized only near
surfaces or interfaces at very low
temperatures.\cite{Matsu2,Fogelstrom,SigristProgress}
Among the heavy fermion superconductors there are two well-known
systems which have $ {\cal T} $-violating bulk superconducting phases: 
UPt$_3$ and U$_{1-x}$Th$_x$Be$_{13}$ ($ 0.017 < x < 0.45 $).
These materials show superconducting double transitions, and
$ {\cal T} $-violation is associated with the second of the two
transitions.\cite{HF-REVIEW}
A more recent candidate for $ {\cal T} $-violating superconductivity
is Sr$_2$RuO$_4$,\cite{Maeno} a layered perovskite compound.
Experimentally, Sr$_2$RuO$_4$ shows clear features of a strongly
correlated quasi-two-dimensional Fermi liquid above the onset
temperature of superconductivity $T_c=$1.5\,K. 
It was suggested\cite{RSB} that, since this system in the normal state 
behaves like a two-dimensional analogue to $^3$He, the superconducting
phase would correspond to the superfluid A-phase,
which is a well-known $ {\cal T} $-violating pairing
state.\cite{Leggett,Volovik1,Volovik2} 
Indeed, the experimental evidence\cite{Luke,Ishida} for unconventional
superconductivity in Sr$_2$RuO$_4$
is consistent with the basic order parameter
symmetry,\cite{Luke,Ishida,PhysicaC}
\begin{equation}
\bbox{d} (\bbox{k}) =
 \hat{\bbox{z}}
 \Delta
 \frac{k_x \pm i k_y}{k_F},
\label{chirp}
\end{equation}
a $p$-wave (spin-triplet) state.
Here we have used the standard notation of the $\bbox{d}$-vector to
represent the order parameter
of the triplet state:\cite{Leggett}
$ \widehat{\Delta}(\bbox{k}) =
 i \bbox{d}(\bbox{k}) \cdot \bbox{\sigma} \sigma^y $,
where $\sigma^i$ are the Pauli matrices.
The $\bbox{d}$-vector (\ref{chirp}) pointing along the $z$ direction 
implies that the spin part of the Cooper pair wave function
is the spin-triplet state with $ S_z = 0 $, i.e.,
in-plane equal-spin pairing
(the $z$ direction is along the $c$-axis of Sr$_2$RuO$_4$).
In a system with cylindrical symmetry the orbital part of the pair
wave function is a state with finite angular momentum along the
$z$-axis, $ L_z = \pm 1 $.
Obviously the state represented by Eq.\ (\ref{chirp}) is
``ferromagnetic,'' and is also called {\it chiral $p$-wave} state. 

The chirality of this superconducting state is characterized by a
topological number $ N $ defined by\cite{Volovik1,Volovik2}
\begin{equation}
N  =
\frac{1}{4\pi}
\int\limits^\infty_{-\infty}dk_x
\int\limits^\infty_{-\infty}dk_y
\widehat{\bbox{m}}\cdot
\left(
 \frac{\partial\widehat{\bbox{m}}}{\partial k_x}
 \times
 \frac{\partial\widehat{\bbox{m}}}{\partial k_y}
\right).
\label{topologicalN}
\end{equation}
The unit vector
$\widehat{\bbox{m}}(k_x,k_y)$ is
\begin{equation}
\widehat{\bbox{m}}
=\frac{\bbox{m}}{|\bbox{m}|},\qquad
\bbox{m}=
\left({\rm Re}\,d_z, {\rm Im}\,d_z,
 \epsilon_{\bbox{k}}\right),
\end{equation}
where $ \epsilon_{\bbox{k}}=(k_x^2+k_y^2)/2m-\mu $
is the kinetic energy measured from the chemical potential $\mu$.
(Throughout this paper we will assume the cylindrical
symmetry whenever microscopic modeling is necessary, although the
inclusion of crystal field effects is straightforward.
Note that, being topological, the number $N$ can be defined
without resorting to the cylindrical symmetry.) 
The topological number $N$ corresponds to the multiplicity of the
wrapping of a sphere $S^2$ in the mapping of
${\bf R}^2\cup\{\infty\}\simeq S^2$ to another $S^2$ representing the
unit vector $\widehat{\bbox{m}}$.
The vector $\bbox{m}$ does not vanish at any
$\bbox{k}$, allowing us to define the topological number $N$ uniquely.
In the chiral $p$-wave state (\ref{chirp}) the topological number $N$
is either $ +1  $ or $ -1 $ depending on
the chirality of the state or the orbital angular momentum. 

A chiral superconductor has gapless chiral edge modes
at interfaces between states of opposite chiralities (domain walls) or 
between a chiral state and a vacuum (surface), as
dictated by the index theorem.
Volovik pointed out\cite{Volovik2} that the topological invariant $N$
determines the number of gapless edge
modes per spin at a boundary between different states.\cite{Volovik2}
This yields $ 2 |N_1 - N_2| $ such modes at the interface of two
domains characterized by the topological numbers $ N_1 $ and $ N_2 $,
respectively. 
At an interface between a chiral $p$-wave state and a vacuum
we find two modes (one per spin).
These states can be easily understood in terms of
Andreev bound states.\cite{Buch,Yamashiro,Honerkamp,Matsu1} 
Recent point contact experiments provide
the first indication for these states at the surface of
Sr$_2$RuO$_4$.\cite{LAUBE} 

The presence of gapless chiral edge modes is reminiscent of quantum
Hall fluids.
One can thus expect a chiral superconductor to have a Hall effect.
This is a spontaneous Hall effect (SHE) in the sense that a transverse
voltage appears in response to an external current even without
an external magnetic field applied to the superconductor.\cite{Volovik1}
The Hall conductance is not quantized, however, contrary to a naive
analogy.
This is because the electric current can be carried as a supercurrent
by superconducting condensate, or, in other words, the local charge 
density is not a conserved quantity in a
superconductor.\cite{Goryo,Senthil,Read}
In fact, it has been shown\cite{Yakovenko,Senthil,Read} that the 
spin and the thermal Hall conductance are quantized to the topological 
number $N$ in appropriate units.
Even though it is non-universal, the spontaneous (charge) Hall effect
is interesting in its own right because it is an effect which can be 
experimentally measured, in principle. Thus we would like to
investigate in detail in this paper the origin and magnitude of the
effect.
To this end, it is necessary to understand what one would actually
measure in an experiment trying to observe the SHE.
Computing a diagram for the current-current correlation function in a
bulk superconductor does not give a proper solution to this problem.
On the contrary, we have to understand physics near a surface of a
chiral superconductor, taking into account issues such as
(1) the spatial variation of the order parameter near the surface,
(2) the spontaneous current running along the surface of the chiral
superconductor, and (3) screening effects for both magnetic field and
electric field in the superconductor.
We study the surface properties using three different approaches:
a numerical self-consistent analysis of the Bogoliubov-de Gennes (BdG)
equation, an extended Ginzburg-Landau (GL) theory with a scalar
potential, and a phenomenological hydrodynamic formulation.
We find that there are two basic contributions to the SHE.
The first is the Bernoulli force due to spontaneous
currents which flow near the surface.
They are the consequence of the chiral surface states, but can also be 
understood in terms of a transverse order parameter textures.
Naturally they lead to a Hall response.
The second originates from the intrinsic magnetic moment of the Cooper
pairs in the chiral $p$-wave state.
The former contribution is present in both chiral and non-chiral
superconductors, since both can have spontaneous surface currents,
while the latter contribution exists only in chiral superconductors.
We therefore conclude that even a non-chiral (antiferromagnetic) 
$ {\cal T} $-violating superconductor can have a SHE.

The organization of this paper is as follows.
In Sec.~II we introduce a simple microscopic model for the chiral
$p$-wave superconductor and solve the BdG equation self-consistently
at zero temperature.
In Sec.~III we present the extended GL theory and derive basic
equations.
They are used in Sec.~IV to analyze the SHE near $T_c$.
In Sec.~V we present parameters appearing in the extended GL theory
obtained from a microscopic weak-coupling analysis.
The GL theory is the basis to formulate an effective hydrodynamical
theory presented in Sec.~VI, which allows us to interpret and
estimate the SHE in a simple way distinguishing the two
contributions mentioned above.
In Sec.~VII we also demonstrate the SHE in the non-chiral $s+id$-wave
state by the numerical BdG approach. 
The results are summarized in Sec.~VIII.

\section{Self-consistent Bogoliubov-de Gennes Analysis}

\subsection{Model formulation}

We consider a chiral $p$-wave superconductor in which
energy band and order parameter have no momentum dependence in the $z$ 
direction.
We may thus treat the system as if it is two-dimensional.
The starting point of our analysis is a mean-field Hamiltonian for a
two-dimensional spin-triplet superconductor with
$\bbox{d}\parallel\hat{\bbox{z}}$.
The Hamiltonian is given by
$H_{\rm MF}=\int dx\int dy\,{\cal H}_{\rm MF}$ with the Hamiltonian
density
\begin{eqnarray}
{\cal H}_{\rm MF}&=&
\sum_\sigma
\psi^\dagger_\sigma(\bbox{r})h_0(\bbox{r})\psi_\sigma(\bbox{r})
+\frac{1}{g}|\bbox{\eta}(\bbox{r})|^2\nonumber\\
&&
-\frac{i}{2k_F}
\biggl\{
\bbox{\eta}(\bbox{r})\cdot\left[
\psi^\dagger_\downarrow(\bbox{r})\nabla\psi^\dagger_\uparrow(\bbox{r})
+\psi^\dagger_\uparrow(\bbox{r})\nabla\psi^\dagger_\downarrow(\bbox{r})
\right]
+\bbox{\eta}^*(\bbox{r})\cdot\left[
\psi_\downarrow(\bbox{r})\nabla\psi_\uparrow(\bbox{r})
+\psi_\uparrow(\bbox{r})\nabla\psi_\downarrow(\bbox{r})
\right]
\biggr\},
\label{MFHamiltonian}\\
h_0(\bbox{r})&=&
\frac{1}{2m}
\left[
-i\hbar\nabla+\frac{e}{c}\bbox{A}(\bbox{r})
\right]^2
 -\mu-eA_0(\bbox{r}),
\label{h_0}
\end{eqnarray}
where $\psi_\sigma$ is the annihilation operator of electron with spin 
$\sigma=\uparrow,\downarrow$,
$g$ is the coupling constant of the attractive interaction that is
responsible for $p$-wave pairing ($g>0$),
and $\nabla=(\partial_x,\partial_y)$.
The superconducting order parameter should satisfy the
(self-consistence) gap equation,
obtained from $(\delta/\delta\bbox{\eta}^*)\langle H_{\rm MF}\rangle=0$,
\begin{equation}
\bbox{\eta}=(\eta_x,\eta_y)=
\frac{ig}{2k_F}
\left\langle\psi_\uparrow(\bbox{r})[\nabla\psi_\downarrow(\bbox{r})]
- [\nabla\psi_\uparrow(\bbox{r})] \psi_\downarrow(\bbox{r})\right\rangle.
\label{selfconsistent}
\end{equation}
We are interested in systems with boundaries in which the order
parameter $\bbox{\eta}$ depends on
$\bbox{r}=(x,y)$.
The uniform system with $\bbox{d}$ given in
Eq.~(\ref{chirp}) corresponds to the case
$\bbox{\eta}=\Delta(1,\pm i)$.

The equations of motion for the field $\psi_\sigma$ are
$i\hbar\partial_t\psi_\sigma=[\psi_\sigma,H_{\rm MF}]$.
Substituting
$\psi_\uparrow=ue^{-i\epsilon t/\hbar}$ and
$\psi^\dagger_\downarrow=ve^{-i\epsilon t/\hbar}$ into them,
we arrive at the BdG equation:
\begin{mathletters}
\begin{eqnarray}
&&
h_0(\bbox{r})u_n(\bbox{r})
-\frac{i}{2k_F}
 [\bbox{\eta}\cdot\nabla v_n+\nabla\cdot(v_n\bbox{\eta})]
= \epsilon_n u_n(\bbox{r}), \\
-&&{h_0}^*(\bbox{r}) v_n(\bbox{r})
-\frac{i}{2k_F}
 [\bbox{\eta}^*\cdot\nabla u_n+\nabla\cdot(u_n\bbox{\eta}^*)]
= \epsilon_n v_n(\bbox{r}).
\end{eqnarray}
\label{BdG}
\end{mathletters}\noindent
The wave functions are normalized in two dimensions:
$\int dx\int dy\left(|u_n|^2+|v_n|^2\right)=1$.
When $(u_n, v_n)$ is an eigenfunction with energy $\epsilon_n$,
$(v_n^*, u_n^*)$ is an eigenstate with energy $-\epsilon_n$.
With these eigenfunctions the field operators are expanded as
\begin{equation}
\pmatrix{\psi_\uparrow(\bbox{r},t)\cr
         \psi_\downarrow^\dagger(\bbox{r},t)\cr}
=\sum_{\epsilon_n \geq 0}
\pmatrix{u_n & v_n^* \cr v_n & u_n^* \cr}
\pmatrix{
\gamma_{n\uparrow}e^{-i\epsilon_nt/\hbar}\cr
\gamma_{n\downarrow}^\dagger e^{i\epsilon_nt/\hbar}\cr}.
\label{expansion}
\end{equation}
The ground state $|0\rangle$ satisfies $\gamma_{n\sigma}|0\rangle=0$.

Gapless chiral edge modes are present at
boundaries of a chiral $p$-wave superconductor.
To illustrate this within our model, it is sufficient to solve the BdG
equation with a simplified gap function with step-function form,
$\bbox{\eta}=\Theta(x)\Delta(1,i\varepsilon)$, where $\Theta(x)$ is
the Heaviside step function and $\varepsilon=\pm1$ is the chirality of
the condensate.
The self-consistent solution of the BdG equation will
be presented in the next subsection. 
The bound state solution to Eq.~(\ref{BdG}) with energy eigenvalue 
$\epsilon=\Delta\sin\theta$ ($-\pi/2<\theta<\pi/2$) and with boundary
condition $u_n=v_n=0$ at $x=0$ is given by
\begin{equation}
\pmatrix{u_\theta(\bbox{r}) \cr v_\theta(\bbox{r}) \cr}=
\sqrt{\frac{2}{\xi_0L_y}}
\exp\left(-\frac{x}{\xi_0}+i\varepsilon k_Fy\sin\theta\right)
\sin(k_Fx\cos\theta)
\pmatrix{e^{i\pi/4} \cr e^{-i\pi/4} \cr},
\label{bound state}
\end{equation}
where $L_y$ is a linear scale of the superconductor in $y$-direction. 
We have used the Andreev approximation valid for $\Delta\ll \mu$.
The amplitude of the edge state decays in the bulk on the length
$\xi_0=\hbar v_F/\Delta$.
With the wave number in the $y$ direction
$k_y=\varepsilon k_F\sin\theta$, the energy dispersion of the chiral
edge mode is $\epsilon(k_y)=\varepsilon\Delta k_y/k_F$.

We now calculate current carried by the chiral edge
mode at zero temperature.
The current density $\bbox{J}$ is defined as
\begin{equation}
J_i=-c\frac{\delta}{\delta A_i}\int d^2r{\cal H}_{\rm MF}
=-e\sum_\sigma
\left\{
\frac{\hbar}{2im}
\left[
\psi^\dagger_\sigma (\partial_i\psi_\sigma)
-(\partial_i\psi^\dagger_\sigma)\psi_\sigma
\right]
+\frac{e}{mc}A_i\psi_\sigma^\dagger\psi_\sigma
\right\},
\end{equation}
where the fields $\psi_\sigma$ are expanded as in (\ref{expansion}).
At this level a finite contribution comes only from the edge mode
(\ref{bound state}).
Thus, we may restrict the summation over the eigenstates to
$0<\theta<\pi/2$.
The total charge current running along the edge is given by
\begin{eqnarray}
I_y&=&\int_0^\infty dx\langle0| J_y |0\rangle
=-\frac{e\hbar}{m}\int_0^\infty\!dx\,
{\rm Im}
\langle0|
\psi_\uparrow^\dagger\partial_y\psi_\uparrow
+\psi_\downarrow^\dagger\partial_y\psi_\downarrow
|0\rangle
\nonumber\\
&=&
-\frac{2e\hbar}{m}
\frac{k_FL_y}{2\pi}
\int^1_0\!d(\sin\theta) \int^\infty_0\!dx\,
{\rm Im}\left(v_\theta\partial_y v_\theta^*\right)
=\frac{\varepsilon e\mu}{2\pi\hbar},
\label{Andreev-current}
\end{eqnarray}
where we have set $\bbox{A}=0$, ignoring the vector potential induced
by the spontaneous current, i.e., diamagnetic screening currents.
The current $I_y$ is spontaneously running along the edge of the
chiral $p$-wave superconductor.
If the chemical potential could be shifted by a constant as
$\mu\to\mu+eV$ in some way, then the spontaneous current would change by
$\varepsilon e^2V/h$.
From this simple-minded argument it is tempting to
conclude a universal value of the Hall conductance.
This argumentation is invalid, because both the
superconducting condensate and the edge states carry current
and, furthermore, the constant shift of the chemical potential is not
realistic to describe a Hall measurement.
Indeed a careful analysis of the physics of the superconductor surface
region is necessary as we will demonstrate below.

\subsection{Solution for Hall bar geometry}

In this section we study the transverse voltage induced by an
externally driven current by solving the BdG equations (\ref{BdG})
self-consistently for a system with Hall bar geometry.
We model the Hall bar by a two-dimensional system of width $L_x$ in 
the $x$ direction and length $L_y$ in the $y$ direction. 
The currents run along the $y$ direction, in which we impose
the periodic boundary conditions.
The superconducting state with the
symmetry of the chiral $p$-wave state is
parameterized by $\bbox{\eta}=(\Delta_x,i\Delta_y)$, where $\Delta_x$
and $\Delta_y$ are real functions of $\bbox{r}$.
The calculation is done for zero temperature. 

In the Hall bar geometry it is convenient to use the following two
basis sets of wave functions,
\begin{equation}
\left(
  \matrix{
    \psi_{\rm sin}(\bbox{k},\bbox{r}) \cr
    \psi_{\rm cos}(\bbox{k},\bbox{r}) \cr
  }
\right)
=\sqrt{2 \over L_x L_y}e^{ik_y y}
\left(
  \matrix{
    \sin{k_x x} \cr
    \cos{k_x x} \cr
  }
\right),
\end{equation}
where $k_y=2l\pi/L_y$ and $k_x=m\pi/L_x$ ($l, m$ : integer) by setting
either the wave function or its first derivative to zero at $ x=0 $
and $ L_x $.
We solve the problem twice: (i) with the Dirichlet boundary condition 
at $x=0$ and $L_x$, and (ii) with the Neumann boundary condition.
Afterwards we take the average of both solutions in order to remove
unphysically rapid changes of the electron 
density at the surface. 
Expanding $u_n(\bbox{r})$ and $v_n(\bbox{r})$ as
\begin{equation}
\pmatrix{
    u_n(\bbox{r}) \cr
    v_n(\bbox{r}) \cr
  }
=
\sum_{\bbox{k}}
\pmatrix{
    \tilde{u}_n(\bbox{k}) \cr
    \tilde{v}_n(\bbox{k}) \cr
  }
\psi_{\sin,\cos}(\bbox{k},\bbox{r}),
\end{equation}
we can solve the gap equation numerically by diagonalization.
The order parameter is determined self-consistently from the
gap equation (\ref{selfconsistent}) for $T=0$
\begin{equation}
(\Delta_x,i\Delta_y)=
\frac{ig}{2k_F}
\sum_{E_n>0} \left[
u_n(\bbox{r})\nabla{v_n}^*(\bbox{r})
-v_n^*(\bbox{r})\nabla u_n(\bbox{r})
\right],
\label{eqn:gapequation}
\end{equation}
where the coupling constant $g$ is chosen to give 
$\Delta_x=\Delta_y=\Delta_0$ in a bulk chiral $p$-wave
superconductor. 
That is, $g$ is obtained by solving the gap equation:
\begin{equation}
\frac{1}{g}=
\frac{1}{(4\pi k_F)^2}\int d^2k
\frac{k^2}{\sqrt{\epsilon_k^2 + (\Delta_0 k/k_F)^2}}
=\frac{m}{8\pi\hbar^2}\int^{\omega_c}_{-\omega_c}d\epsilon
\frac{1+\epsilon/\mu}{\sqrt{\epsilon^2+\Delta^2_0(1+\epsilon/\mu)}}
=\frac{m}{8\pi\hbar^2}(I_1+I_2),
\end{equation}
where $\omega_c$ is a cutoff energy and
\begin{eqnarray*}
I_1&=&\int_{-\omega_c}^{\omega_c}d\epsilon
{1 \over \sqrt{\epsilon^2+{\Delta_0}^2(1+\epsilon/\mu)}}
=
\ln\left|
{
2\omega_c+{\Delta_0}^2/\mu
+2\sqrt{\omega_c^2+{\Delta_0}^2(1+\omega_c/\mu)}
 \over
-2\omega_c+{\Delta_0}^2/\mu
+2\sqrt{\omega_c^2+{\Delta_0}^2(1-\omega_c/\mu)}
}
\right|,\\
I_2&=&
{1 \over \mu}\int_{-\omega_c}^{\omega_c}d\epsilon
{\epsilon \over\sqrt{\epsilon^2+{\Delta_0}^2(1+\epsilon/\mu)}}
=
\frac{1}{\mu}\left[
\sqrt{\omega_c^2+{\Delta_0}^2 \left(1+\frac{\omega_c}{\mu}\right)}
-\sqrt{\omega_c^2+{\Delta_0}^2 \left(1-\frac{\omega_c}{\mu}\right)}
\right]
-{{\Delta_0}^2 I_1 \over 2\mu^2}.
\end{eqnarray*}

The solutions to the BdG equation (\ref{BdG}) determine the electron
density and current density:
\begin{mathletters}
\begin{eqnarray}
n_{\rm e}(\bbox{r})&=&
2\sum_{E_n>0}|v_n(\bbox{r})|^2,
\label{eqn:charge}\\
J_y(\bbox{r})&=&
{i e\hbar \over m}\sum_{E_n>0}
\left[
  v_n(\bbox{r})
    \frac{\partial v_n^*(\bbox{r})}{\partial y}
- v_n^*(\bbox{r})
    \frac{\partial v_n(\bbox{r})}{\partial y}
\right]
-{e^2 \over m c}n_{\rm e}(\bbox{r})
 A_y(\bbox{r}),
\label{eqn:current}
\end{eqnarray}
\end{mathletters}
where only the $y$ component of the current is non-vanishing in the
Hall bar geometry.
Note that these densities are defined in two dimensions because of
the normalization condition imposed on $u_n$ and $v_n$.
The scalar and vector potentials obey the Maxwell equations,
\begin{equation}
\nabla^2 A_0(\bbox{r})
= - \frac{4\pi e}{d}
\left[n_{\rm b}-n_{\rm e}(\bbox{r})\right],
\qquad
\nabla^2 A_y(\bbox{r})=
-\frac{4\pi}{cd} J_y(\bbox{r}).
\label{Maxwells}
\end{equation}
We have introduced the length $d$ to convert the area densities into
the volume densities.
Physically $d$ corresponds to the spacing between two-dimensional
layers in Sr$_2$RuO$_4$.
To keep overall charge neutrality, we introduce $n_{\rm b}$ as the
density of the uniform positive background charge (jellium model).
The externally injected current $I$ fixes the boundary conditions 
for the vector potential:
\begin{equation}
\left.\frac{\partial A_y}{\partial x} \right|_{x =0}
= - \left.\frac{\partial A_y}{\partial x}\right|_{x =L_x}
= \frac{ 2\pi}{cd} I .
\end{equation}
The self-consistent order parameters, scalar and vector
potentials can be obtained numerically by solving (\ref{BdG}),
(\ref{eqn:gapequation}) and (\ref{Maxwells}) iteratively until
convergence is reached. 
In the iteration step we fix the total number of electrons to the
normal state value by adjusting $\mu$.
In the self-consistent solution electric charge is screened on the
Thomas-Fermi length scale,
$\ell = (\hbar^2 d/4e^2 m)^{1/2}$.
Since we have three material-dependent parameters, $\Delta_0$, $k_F$,
and $d$, we have freedom to change three dimensionless
parameters, $k_F\xi_0$, $\lambda/\xi_0$, and $\ell/\xi_0$.

\subsection{Discussion of the self-consistent solution}

\begin{figure}[t]
\begin{center}
\begin{minipage}[t]{8cm}
\epsfxsize=8cm
\epsfbox{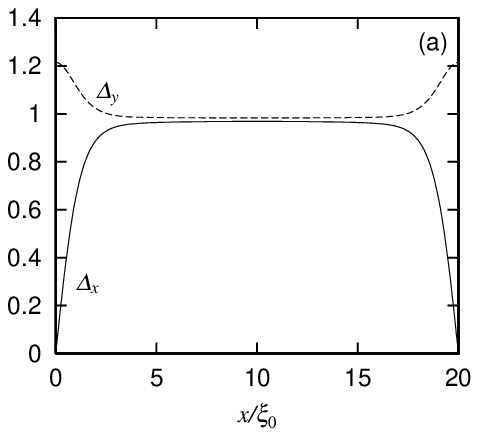}
\end{minipage}
\begin{minipage}[t]{8cm}
\epsfxsize=8cm
\epsfbox{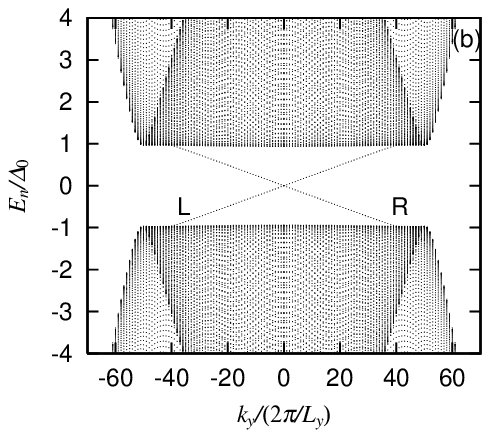}
\end{minipage}
\begin{minipage}[t]{8cm}
\epsfxsize=8cm
\epsfbox{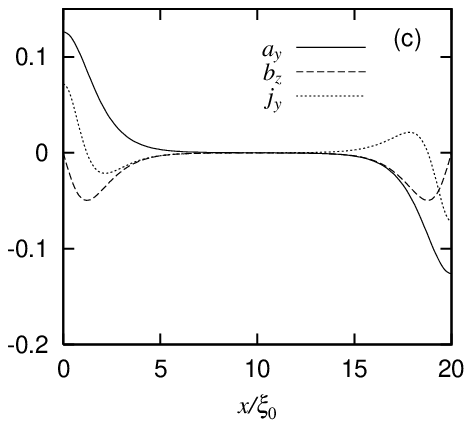}
\end{minipage}
\begin{minipage}[t]{8cm}
\epsfxsize=8cm
\epsfbox{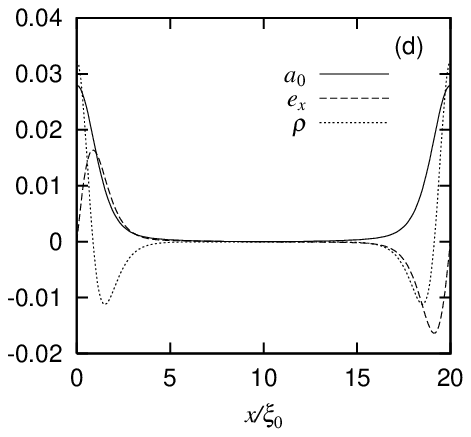}
\end{minipage}
\end{center}
\caption{Self-consistent solution of the Bogoliubov-de Gennes equation
of the $p_x+i p_y$-wave state at $T=0$.
The set of parameters are chosen as
$\xi_0=\hbar v_F/\Delta_0$,
$k_F\xi_0=16$,
$\omega_c=8\Delta_0$,
$\kappa=\lambda/\xi_0=1$,
$ \ell=0.25\xi_0$ and $L_x=L_y=20\xi_0$.
(a) Order parameter scaled by $\Delta_0$,
which is the magnitude of the order parameter in the bulk region.
(b) Energy eigenvalues obtained for sine basis functions.
L and R denote the surface bound states localized near $x=0$ and
$ x = L_x$, respectively.
States with $|E_n| > \Delta_0 $ are extended.
Unit of $k_y$ is $2\pi/L_y$.
(c) Dimensionless vector potential $a_y=(e\xi_0/\hbar c)A_y$,
magnetic field $b_z=\xi_0\partial_x a_y$
and current density $j_y=-\xi_0\partial_x b_z$.
(d) Dimensionless scalar potential $a_0=eA_0/\Delta_0$,
electric field $e_x=-\xi_0\partial_x a_0$ and charge
density $\rho=\xi_0\partial_x e_x$.
}
\label{fig:BdG1}
\end{figure}

The solutions of the BdG equation reveal that the relevant physics of
the Hall bar indeed happens at the two surfaces.
First we consider the solution for the case
where total current along the $y$ direction is zero. 
The order parameter varies strongly at the surface:
$\Delta_x$ is suppressed while $\Delta_y$ rises slightly as shown
in Fig.~\ref{fig:BdG1}(a).
This behavior is connected with the reflection properties of Cooper pairs
at the surface.
Resulting interference effects are destructive for $\Delta_x$,
since the order parameter is odd under reflection at a
surface normal to the $x$ direction.
Note that this order parameter variation is specific for a specular
reflection at the surface and would look different for the case of 
diffuse scattering.
We do not, however, consider this aspect here further. 

At the surface the chiral edge states appear
with a linear dispersion around the Fermi energy.
These are Andreev bound states as a direct consequence of the
chirality of the superconducting state.
The two branches seen in Fig.~\ref{fig:BdG1}(b) belong each to one of
the two edges of the bar.
These chiral edge states generate spontaneous currents
at the surface flowing along $ \bbox{n} \times \hat{{\bf z}} $
($ \bbox{n} $: surface normal vector).
The currents on the two edges of the Hall bar run in opposite
directions, thereby no net current flowing in the Hall bar.
The edge currents generate a magnetic field, which is screened
by counter currents in the interior of the superconductor
[Fig.~\ref{fig:BdG1}(c)].
The length scale of the surface current is the coherence length
$\xi_0$ while the screening currents spread over the London
penetration depth $\lambda$.
The surface magnetic fields generate a finite magnetization 
whose sign depends on the sign of Cooper pair angular momentum 
and the sign of the charge, i.e., electron-like or hole-like Fermi
surface. 

Turning now to the question of the scalar potential and the charge
distribution in the superconductor, we find again that all
interesting features show up only in the surface region.
There is a finite excess charge at the surface, which is screened due
to Thomas-Fermi screening.
As a result the charge density forms a dipole layer and is overall
charge neutral.
This is the constraint that we have imposed by fixing the external
electrical field to zero (note that in our Hall bar geometry a finite
charge or equivalently a finite external electric field would
correspond to an infinite field energy, since the capacitance is
zero).
The dipole layer induces a local electric field and causes a shift of
the scalar potential relative to its
value in the bulk of the superconductor which we choose to be zero.
The charge distributions and potential at the surface are the same on
both sides of the bar. 

Under the assumption that the Hall bar is symmetric about $x=L_x/2$,
there is no potential difference between the two sides.
We would like to mention that scalar potential variations close to the
surface are not unique to chiral superconducting states,
but occur in any superconductor whose order parameter
is influenced by surface scattering. 

\begin{figure}[t]
\begin{center}
\begin{minipage}[t]{8cm}
\epsfxsize=8cm
\epsfbox{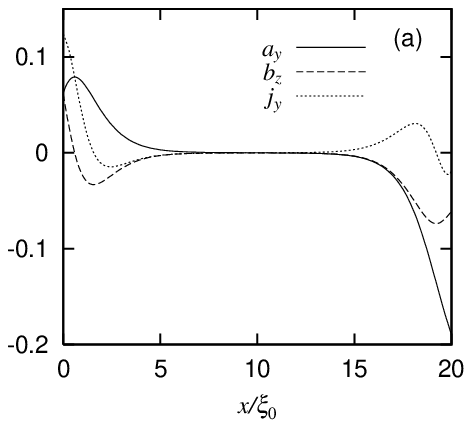}
\end{minipage}
\begin{minipage}[t]{8cm}
\epsfxsize=8cm
\epsfbox{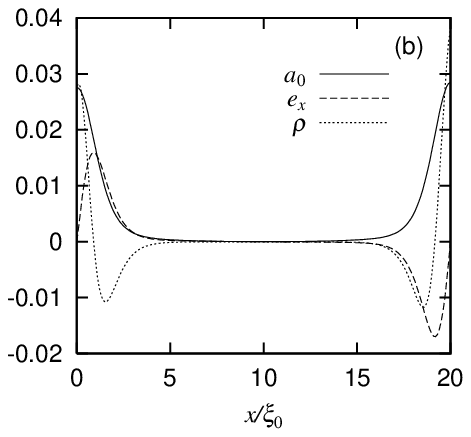}
\end{minipage}
\end{center}
\caption{Self-consistent solution of the Bogoliubov-de Gennes equation
at $T=0$ when current $I=5e\Delta_0/\hbar k_F\xi_0$ is externally
supplied.
Set of parameters are the same as used in Fig. \ref{fig:BdG1}.
(a) Dimensionless vector potential, magnetic field and current density.
(b) Dimensionless scalar potential, electric field and charge density.
}
\label{fig:BdG2}
\end{figure}

If we introduce a net current driven from an external source, this 
external current will be distributed equally to both surfaces
[see Fig.~\ref{fig:BdG2}(a)]. 
This affects the surface states differently for the two sides,
because on one side the external current flows with the spontaneous
current, on the other against it. 
Furthermore, the scalar potential is not equal anymore at the
two sides because the charge dipole layers are modified differently,
as we can see in Fig.~\ref{fig:BdG2}(b).
This transverse voltage difference depends on the orientation of the
external current and appears in the absence of an external magnetic
field, thereby referred to as the SHE.

\begin{figure}[t]
\begin{center}
\begin{minipage}[t]{8cm}
\epsfxsize=8cm
\epsfbox{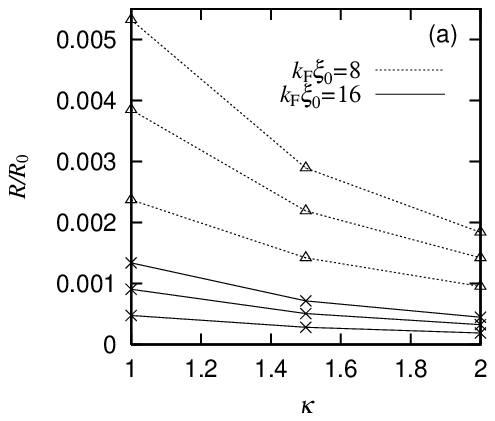}
\end{minipage}
\begin{minipage}[t]{8cm}
\epsfxsize=8cm
\epsfbox{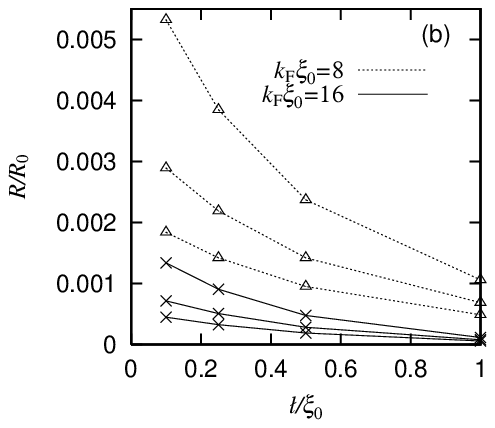}
\end{minipage}
\end{center}
\caption{(a) $\kappa$ dependence of the Hall resistance.
Within the same $k_F\xi_0$,
the three lines from the top correspond to $\ell=0.1\xi_0$,
$\ell=0.25\xi_0$ and $\ell=0.5\xi_0$, respectively.  
(b) $\ell$ dependence of the Hall resistance.
Within the same $k_F\xi_0$,
three lines from the top correspond to $\kappa=1$, $\kappa=1.5$ and
$\kappa=2$, respectively. 
In both figures the Hall resistance $R$ is scaled by $R_0=h/2e^2$.
}
\label{fig:BdG3}
\end{figure}

Our calculation clearly shows a linear relation between the source
current $I$ and the transverse voltage $V_H$ as expected for the Hall
effect.
Deviations occur only when the current approaches the critical value
where the order parameter starts to be strongly affected by the
current. 
In Fig.~\ref{fig:BdG3} we show the $\kappa$, $\ell$ and
$k_F\xi_0$ dependences of the Hall resistance, $ R_H = V_H /I $. 
What is immediately obvious is that the Hall resistance is strongly
suppressed from the quantum unit of resistance $ R_0 = h/2e^2 $.
There is also a strong dependence on the material dependent
parameters, indicating that the behavior is non-universal.
Unfortunately, in the numerical BdG scheme we are limited in the
choice of $ \ell $, $\xi_0$, $\lambda$ and $k_F^{-1}$, because large
difference in their magnitudes demands large-scale computation.
In the next section we study the SHE using the
extended GL theory which will allow us to calculate $R_H$ analytically 
for temperatures close to $T_c$.  
We can already here confirm that the quantitative comparison between
the two methods work very well.
It will also become clear that there are a few contributions
to the transverse voltage.

\section{Ginzburg-Landau formulation}

We formulate an extended Ginzburg-Landau theory based on symmetry
arguments which includes the scalar and vector potentials in a general
form.
This allows us to analyze the anomalous coupling between charge and
magnetic degrees of freedom in chiral superconductors.

\subsection{Ginzburg-Landau free energy}

The paring symmetry of the chiral $p$-wave superconducting state is
characterized by
$\bbox{d}(\bbox{k})
= \hat{\bbox{z}}\Delta (k_x \pm i k_y)/k_F $,
which is a combination of the two degenerate $p$-wave components with
$ p_x $- and $ p_y $-symmetry.
This degeneracy is not lifted when we introduce a tetragonal crystal
field, although the details of the $k$-dependence of $\bbox{d}$ may
change.
The two-fold degeneracy requires that we introduce two
complex order parameter components,
$ \bbox{\eta} = (\eta_x , \eta_y ) $, such that the $\bbox{d}$-vector
becomes
$ \bbox{d}(\bbox{k})
 = \hat{\bbox{z}}(\bbox{\eta} \cdot \bbox{k})/k_F $.
The free energy has to be a scalar under the transformations of the
symmetry group
\begin{equation}
{\cal G} = D_{4h} \times {\cal T} \times U(1),
\end{equation}
where the tetragonal point group $ D_{4h} $ includes the simultaneous
transformation of orbital and spin degree of freedom as a consequence
of spin-orbit coupling, $ {\cal T} $ denotes the time-reversal
symmetry and $ U(1) $ the gauge symmetry.
The GL free energy expansion for $\bbox{\eta}$ with the symmetry
$ {\cal G} $ is well known:\cite{SigristUeda} $F=\int d^3r {\cal F}$,
where 
\begin{eqnarray}
{\cal F} =
&&
a \bbox{\eta}\cdot\bbox{\eta}^*
+ b_1 \left(\bbox{\eta}\cdot\bbox{\eta}^*\right)^2
+ \frac{b_2}{2}\left(\eta^{*2}_x \eta^2_y
                   + \eta^2_x \eta^{*2}_y\right)
+ b_3 |\eta_x|^2 |\eta_y|^2  \nonumber\\
&&
+ K_1 \left( |D_x \eta_x|^2 + |D_y \eta_y|^2 \right)
+ K_2 \left( |D_x \eta_y|^2 + |D_y \eta_x|^2 \right)
 \nonumber \\
&&
+ K_3 \left[ (D_x \eta_x)^* (D_y \eta_y) + {\rm c.c.}\right]
+ K_4 \left[ (D_x \eta_y)^* (D_y \eta_x) + {\rm c.c.}\right]
 \nonumber \\
&&
+ K_5 \left( |D_z \eta_x |^2 + |D_z \eta_y|^2 \right)
+ \frac{(\nabla \times\bbox{A})^2}{8 \pi}.
\label{standardfe}
\end{eqnarray}
The coefficients $a$, $b_i$ and  $K_i$ are non-universal real numbers
that depend on the details of the material.
The coefficient $a$ is negative below $T_c$ ($a\propto T-T_c$).
For the choice $ 0 < b_2 < 4 b_1 + b_3 $ and $ b_3 < b_2 $, we find
the homogeneous phase
$ \bbox{\eta} = \eta_0 (1, \pm i)$ with
$ \eta^2_0 (T) = |a|/( 4 b_1 - b_2 + b_3)$.
The gradient terms are explicitly gauge invariant by the definition
$\bbox{D} = \nabla + i (2e/\hbar c)\bbox{A}$.
Equation (\ref{standardfe}) is the standard free energy density used
to study the response to the magnetic field with
the Coulomb gauge $\nabla\cdot\bbox{A}=0$.

As is well known,\cite{Samokhin} the intrinsic orbital
angular momentum in a chiral $p$-wave state is related to the
difference of the two terms proportional to $K_3$ and $K_4$:
\begin{equation}
(D_x\eta_x)^*D_y\eta_y-(D_x\eta_y)^*D_y\eta_x+{\rm c.c.}
=
\left[
\nabla\times\left(\eta^*_x\bbox{D}\eta_y-\eta^*_y\bbox{D}\eta_x\right)
+i\frac{2e}{\hbar c}(\eta^*_y\eta_x-\eta^*_x\eta_y)
 \nabla\times\bbox{A}
\right]_z ,
\label{K_3-K_4}
\end{equation}
where it is understood that the $z$ component of a three-dimensional
vector  is kept in the right hand side.
The integral of the first term gives a surface term which we may
discard, while the second term represents the Zeeman energy of a
magnetic moment coming from the intrinsic angular moment
($\propto i\bbox{\eta}\times\bbox{\eta}^*$).

For our purpose it is essential to
include additional terms that are coupled to either the scalar
potential $ A_0 $ or
the electric field $ \bbox{E} = - \nabla A_0 $.
We consider the stationary situation and fix the gauge so that 
$ A_0 $ is zero in the undisturbed homogeneous superconductor. 
It is more convenient to use the Lagrangian formulation,
in which the Maxwell equations simply follow from variation
with respect to $ A_\mu $.
The Lagrangian density involving the scalar potential is given by
\begin{eqnarray}
{\cal L}_{e} =
&&
\widetilde{K}_1 A_0 (|D_x \eta_x|^2 + |D_y \eta_y|^2)
+ \widetilde{K}_2 A_0(|D_x \eta_y|^2 + |D_y \eta_x|^2) \nonumber\\
&&
+ \widetilde{K}_3 A_0 [(D_x \eta_x)^* (D_y \eta_y)
                     + (D_x \eta_x) (D_y \eta_y)^* ] \nonumber \\
&& 
+ \widetilde{K}_4 A_0 [(D_x \eta_y)^* (D_y \eta_x)
                     + (D_x \eta_y) (D_y \eta_x)^* ]
+ \widetilde{K}_5 A_0 (|D_z \eta_x|^2 + |D_z \eta_y|^2) \nonumber\\
&& 
+C_1 [ E_x \eta^*_x (D_x \eta_x)
    + E_y \eta^*_y (D_y \eta_y) + {\rm c.c.}]
+ C_2 [ E_x \eta^*_y (D_x \eta_y)
      + E_y \eta^*_x (D_y \eta_x) + {\rm c.c.}] \nonumber \\
&&
+ C_3 [ E_x \eta^*_x (D_y \eta_y)
      + E_y \eta^*_y (D_x \eta_x) + {\rm c.c.}]
+ C_4 [ E_x \eta^*_y (D_y \eta_x)
      + E_y \eta^*_x (D_x \eta_y) + {\rm c.c.}] \nonumber \\
&&
+ C_5 [E_z \eta^*_x (D_z \eta_x)
     + E_z  \eta^*_y (D_z \eta_y) + {\rm c.c.}]
- \frac{(\nabla A_0)^2}{8 \pi}
- \frac{A^2_0}{8\pi\ell^2} 
\label{edfe}
\end{eqnarray}
with $ \widetilde{K}_i, C_i $ and $ \ell $ being real numbers.
The coefficients will be derived for the weak-coupling
limit in Sec.~V.
In this form it is easy to verify that each term is individually
invariant.
The term $ A^2_0/8\pi\ell^2 $ describes the screening of the
electric charge in the metallic and superconducting state, where
$\ell$ is the Thomas-Fermi screening length.
The choice of this form fixes $ A_0 $ to zero in the bulk of the
superconductors, which corresponds to the chemical potential as
required by our choice of gauge.
We emphasize that ${\cal L}_e+{\cal F}$ plays a role of
the Lagrangian density for $A_\mu$.

We notice that the $\widetilde K_i$ terms are closely related with the
$K_i$ terms. On the other hand, the $C_i$ terms have no relatives in
Eq.~(\ref{standardfe}).
The difference of the $C_3$ and $C_4$ terms contains a contribution
$i(\eta_x\eta_y^*-\eta_y\eta_x^*)
(A_x\partial_yA_0-A_y\partial_xA_0)$.
It describes a coupling between the scalar and the
vector potentials and is similar to the Chern-Simons  (CS) term.
Since the fields are static, however, it is not exactly 
the same as the CS term.
Thus we shall call it a CS-like term.
It can also be interpreted as representing a coupling between the
intrinsic magnetic moment ($\propto i\bbox{\eta}\times\bbox{\eta}^*$)
and the magnetic field in the presence of an electric field.
In other words, the CS-like term describes the reaction of the
intrinsic magnetic moment to a change in the Cooper pair
density.\cite{MINEEV}
The CS-like term proportional to $C_3-C_4$ will play an essential role
in the SHE besides others.

\subsection{Equations for the electromagnetism}

We derive equations describing the electromagnetic properties of the 
superconductor from variation of $\int d^3r({\cal L}_e+{\cal F})$ with
respect to $ A_0 $ and $ \bbox{A} $.
The equation for the scalar potential has the form,
\begin{equation}
 - \nabla^2 A_0 
 + \frac{A_0}{\ell^2}
 = 4 \pi \left[
 \tilde{\rho}
 - \nabla \cdot \left( \bbox{P} + \bbox{\Pi} \right) \right],
\label{Maxwell}
\end{equation}
where
\begin{mathletters}
\begin{eqnarray}
\tilde{\rho} & = & 
\widetilde{K}_1 (|D_x \eta_x|^2 + |D_y \eta_y|^2)
 + \widetilde{K}_2 (|D_x \eta_y|^2 + |D_y \eta_x|^2)
 + \widetilde{K}_3 [(D_x \eta_x)^* (D_y \eta_y)
                  + (D_x \eta_x) (D_y \eta_y)^* ] \nonumber \\
&& 
 + \widetilde{K}_4 [(D_x \eta_y)^* (D_y \eta_x)
                  + (D_x \eta_y) (D_y \eta_x)^* ]
 + \widetilde{K}_5 (|D_z \eta_x|^2 + |D_z \eta_y|^2), \\
&&\nonumber\\
\bbox{P}
& = &
-\pmatrix{
\partial_x (C_1|\eta_x|^2 + C_2|\eta_y|^2)
 + \frac{1}{2}(C_3+C_4)\partial_y
    (\eta_x^*\eta_y+\eta_x\eta_y^*) \cr
\partial_y (C_1|\eta_y|^2 + C_2|\eta_x|^2)
 + \frac{1}{2}(C_3+C_4)\partial_x
    (\eta_x^*\eta_y+\eta_x\eta_y^*) \cr
C_5 \partial_z |\bbox{\eta}|^2}, \\
&&\nonumber\\
\bbox{\Pi} & = &
\frac{1}{2}(C_3 - C_4)
\pmatrix{
\eta_y^*D_y\eta_x+\eta_yD_y^*\eta_x^*
-\eta_x^*D_y\eta_y-\eta_xD_y^*\eta_y^*
\cr
\eta_x^*D_x\eta_y+\eta_xD_x^*\eta_y^*
-\eta_y^*D_x\eta_x-\eta_yD_x^*\eta_x^*
\cr
0\cr}.
\end{eqnarray}
\label{charges}
\end{mathletters}
The two terms on the left hand side of Eq.\ (\ref{Maxwell})
describe the screening of the electric field.
The quantity $\tilde{\rho}$ is the charge density
induced by variations of the order parameter.
The vector density $\bbox{P}$ represents the
electric polarization caused by inhomogeneities of the superconducting
condensate. 
Both $\tilde{\rho}$ and $\bbox{P}$ appear as a
consequence of the fact that the local change in the condensate, both
in the order parameter and in the supercurrent, leads to a
redistribution of the electric charge.
This is most easily seen in $\tilde{\rho}$, which is related to
the gradient term of the free energy.
Note that terms of this kind are also present in conventional
superconductors.
On the other hand, the vector density $\bbox{\Pi}$ is anomalous and
characteristic of chiral superconductors.

For the sake of simplicity we assume that
the relative phase of the order parameter
component is fixed, i.e.,
$ \eta_x = | \eta_x | \exp(i \phi) $ and
$ \eta_y = i \varepsilon | \eta_y | \exp(i \phi )$
with $\varepsilon=\pm1$, where
$\varepsilon$ is the chirality of the $p$-wave order parameter.
This condition is satisfied in the situations we will study below.
In this case we have
$i(\eta_y^*\eta_x-\eta_x^*\eta_y)=2\varepsilon|\eta_x||\eta_y|$.
The vector $ \bbox{\Pi} $ can be written as
\begin{equation}
\bbox{\Pi} =
2 \varepsilon (C_3 - C_4) |\eta_x| |\eta_y|
\left(\nabla \phi + \frac{2e}{\hbar c} \bbox{A}\right)
 \times \hat{{\bbox{z}}}.
\end{equation}
Its divergence is equivalent to a source charge,
\begin{equation}
\rho^{}_{\Pi} =
- \nabla \cdot \bbox{\Pi}
= - \frac{4e}{\hbar c}\varepsilon|\eta_x||\eta_y|B_z(C_3 - C_4)
 - 2 \varepsilon (C_3 - C_4)
   (\nabla |\eta_x| |\eta_y|) \times
   \left(\nabla \phi
         + \frac{2e}{\hbar c}\bbox{A} \right).
\end{equation}
The first term indicates that the magnetic field $B$ induces the
electric charge whose sign depends on the chirality $\varepsilon$. 
The second term is non-vanishing only when the modulus of the order
parameter has spatial variation.
Terms similar to this also exist in $\tilde{\rho}$.
 
The modified London equation is obtained by variation of
$\int d^3r ({\cal L}_e + {\cal F}) $ with respect to
$ \bbox{A} $:
\begin{equation}
\nabla^2 \bbox{A}
+ \frac{4\pi}{c}
\left( \bbox{J} + \widetilde{\bbox{J}} + \bbox{\Upsilon} \right)
= 0,
\label{London}
\end{equation}
where we find three current contributions.
The first two,
\begin{mathletters}
\begin{equation}
\bbox{J}
= - \frac{4e}{\hbar} \pmatrix{
{\rm Im}\left(
K_1 \eta^*_x D_x \eta_x + K_2 \eta^*_y D_x \eta_y
 + K_3 \eta^*_x D_y \eta_y + K_4 \eta^*_y D_y \eta_x
\right) \cr
{\rm Im}\left(
K_1 \eta^*_y D_y \eta_y + K_2 \eta^*_x D_y \eta_x
 + K_3 \eta^*_y D_x \eta_x + K_4 \eta^*_x D_x \eta_y
 \right) \cr
{\rm Im}\left(
K_5 \eta^*_x D_z \eta_x + K_5 \eta^*_y D_z \eta_y
\right) }
\label{J}
\end{equation}
and 
\begin{equation}
\widetilde{\bbox{J}}
= - A_0 \frac{4e}{\hbar} \pmatrix{
{\rm Im}\left(
\widetilde{K}_1 \eta^*_x D_x \eta_x
+ \widetilde{K}_2 \eta^*_y D_x \eta_y
+ \widetilde{K}_3 \eta^*_x D_y \eta_y
+  \widetilde{K}_4 \eta^*_y D_y \eta_x
 \right) \cr
{\rm Im}\left(
\widetilde{K}_1 \eta^*_y D_y \eta_y
+ \widetilde{K}_2 \eta^*_x D_y \eta_x
+ \widetilde{K}_3 \eta^*_y D_x \eta_x
+ \widetilde{K}_4 \eta^*_x D_x \eta_y
 \right) \cr
{\rm Im}\left(
\widetilde{K}_5 \eta^*_x D_z \eta_x + K_5 \eta^*_y D_z \eta_y
\right) },
\end{equation}
are the supercurrents including the screening
currents.
The second one contributes only if the scalar potential
deviates from its bulk value $ A_0=0 $.
The last term 
\begin{equation}
\bbox{\Upsilon}
= i\frac{2e}{\hbar} (C_3 - C_4) (\eta_x\eta_y^*-\eta_x^*\eta_y)
\bbox{E} \times \hat{\bbox{z}}
\label{Upsilon}
\end{equation}
\label{currents}
\end{mathletters}
is the anomalous contribution, where the electric field acts as a
source of supercurrent. 
Note, however, that $\bbox{\Upsilon}$ does not cause
dissipation because
$ \bbox{E} \cdot \bbox{\Upsilon} = 0 $.
Both anomalous components $\bbox{\Pi}$ and $\bbox{\Upsilon}$ are
proportional to the chirality $\varepsilon$ and originate from
the CS-like term. We would like to emphasize here that the presence of 
the anomalous current contribution does not affect the standard flux
quantization. The current $\bbox{\Upsilon}$ quickly drops to zero
inside the superconducting material because the electrical field is
strongly screened. Thus even a hole containing a net charge would not
violate standard laws of flux quantization.\cite{Goryo}

\section{Transverse voltage in the GL description}

Now we shall study the SHE in a Hall bar geometry on the
level of the extended GL theory.
Since we concluded in the BdG study that the relevant physics lies in
the behavior of the superconductor close to the surface whose
influence is exponentially small in the interior, we concentrate only
on one edge and consider the half-space $x \geq 0 $.
The physics of a Hall bar of a given width $L_x$ follows in a simple
way from the results obtained for a single edge.

\subsection{The surface state}

The boundary conditions for the order parameter are chosen assuming
specular surface scattering.
For the given geometry they read
\begin{equation}
\eta_x |_{x=0} = 0 \quad {\rm and} \quad
D_x \eta_y |_{x=0} = 0,
\label{opbc}
\end{equation}
implying that there is
no current running normal through the surface.\cite{SigristUeda}
We need to consider only the
$x$-component of the electric field and the $z$-component of the
magnetic field, both of which are continuous at the surface and are
functions of $x$. 
It can be easily shown that these boundary conditions are compatible
with the gauge invariance of ${\cal F}_e $.
The self-inductance $L$ and the capacitance $C$ of our system are
taken to be infinity and zero, respectively, such that net current and
total charge in the system should vanish, unless they are imposed by
external sources.
This is important for the choice of the boundary conditions for the
gauge fields.

We present an approximate solution to the GL equation that captures
the essential aspects of the problem.
We assume that the system is in a chiral $p$-wave state of a
single domain with the chirality $\varepsilon$.
First we solve the GL equations to determine the spatial dependence of
the order parameter $\bbox{\eta}$.
To this end we ignore the gauge fields $A_\mu$ in the GL equation.
This is allowed because we are looking for a solution in lowest
order in $\tau=(T_c-T)/T_c$, a small parameter in the theory.
The GL equations for $\eta_x = | \eta_x | $ and
$\eta_y=i\varepsilon|\eta_y|$ are
\begin{mathletters}
\begin{eqnarray}
K_1\frac{d^2|\eta_x|}{dx^2}&=&
a|\eta_x|+2b_1|\eta_x|^3+(2b_1-b_2+b_3)|\eta_x||\eta_y|^2,
\label{GL-for-eta_x}\\
K_2\frac{d^2|\eta_y|}{dx^2}&=&
a|\eta_y|+2b_1|\eta_y|^3+(2b_1-b_2+b_3)|\eta_x|^2|\eta_y|.
\label{GL-for-eta_y}
\end{eqnarray}
\end{mathletters}
To simplify the analysis we consider a special situation where the
coefficients $b_i$ in the GL free energy satisfy the relation
\begin{equation}
2 b_1 = b_2 - b_3,
\label{special-relation}
\end{equation}
which, however, is not satisfied for a cylindrical Fermi surface in
the weak-coupling limit (see Sec.~V).
Under this condition Eqs.~(\ref{GL-for-eta_x}) and
(\ref{GL-for-eta_y}) have the solution,
\begin{equation}
\eta_x = \eta_0 \tanh\left(\frac{x}{\xi}\right),
  \quad
\eta_y = i \varepsilon \eta_0
\label{solution-to-GL}
\end{equation}
with $\eta_0^2 = |a|/2 b_1$ and the coherence
length $ \xi^2 = 2 K_1 / |a|$.
Note that $\eta_x$ vanishes linearly at $x=0$ whereas $\eta_y$ stays
constant. 
If the condition (\ref{special-relation}) is not satisfied, $\eta_y$
deviates slightly at $x\lesssim\xi$ from the bulk value $\eta_0$, but
generally does not vanish at $x=0$, as we have seen in the BdG study.
In general the order parameter shows the following feature:
the component of $\bbox{\eta}$ normal to the surface vanishes
linearly while the perpendicular component is only weakly affected.
As the detailed spatial dependence of $\bbox{\eta}$ is not important
for our semi-quantitative discussion, we will use an approximate form
for $\eta_x$, instead of Eq.~(\ref{solution-to-GL}), which allows us
to proceed with analytic calculations more easily,
\begin{equation}
\eta_x = \eta_0 \left(1-e^{-x/\xi}\right).
\label{approximate-eta_x}
\end{equation}
We would like to mention that we 
have solved the coupled GL equations for more general cases by
numerical means to confirm that our approximations work very well both
qualitatively and quantitatively. 

Having determined the profile of the order parameter, we can calculate
the distribution of the charge and the spontaneous current in the
equilibrium state from Eqs.~(\ref{charges}) and (\ref{currents}).
The equations for the scalar and vector potential have the
following form:
\begin{eqnarray}
&&
\frac{d^2A_0}{dx^2}-\frac{A_0}{\ell^2}
-\frac{\varepsilon}{l_1}\left(1-e^{-x/\xi}\right)\frac{dA_y}{dx}
-\frac{\varepsilon A_y}{\xi l_1 l_2}(l_1+l_2)
 e^{-x/\xi}
+\frac{A_y^2}{e l_3(x)}
=\frac{4\pi\eta_0^2}{\xi^2}
 \left[
   2C_1e^{-x/\xi}-\left(4C_1+\widetilde{K}_1\right)e^{-2x/\xi}
 \right],
\label{Maxwell1}\\
&&
\frac{d^2A_y}{dx^2}
-\frac{A_y}{\lambda^2}
-\frac{\varepsilon}{l_1}\left(1-e^{-x/\xi}\right)\frac{dA_0}{dx}
+\frac{\varepsilon A_0}{\xi l_2}e^{-x/\xi}
-\frac{2A_0A_y}{e l_3(x)}
=-\varepsilon  \frac{e}{\xi \lambda l_4} e^{-x/\xi},
\label{Maxwell2}
\end{eqnarray}
where we have introduced parameters $l_i$ $(i=1\sim4)$
that have dimension of length:
\begin{mathletters}
\begin{eqnarray}
\frac{1}{l_1}&=&
\frac{16\pi e\eta_0^2}{\hbar c}(C_3-C_4),
\label{l_1}\\
\frac{1}{l_2}&=&
\frac{16\pi e\eta_0^2}{\hbar c}\widetilde{K}_3,
\label{l_2}\\
\frac{1}{l_3(x)}&=&
\frac{16\pi e^3\eta_0^2}{\hbar^2 c^2}
\left[\widetilde{K}_1+\widetilde{K}_2\left(1-e^{-x/\xi}\right)^2\right],
\label{l_3}\\
\frac{1}{l_4}&=&
\frac{16\pi\eta_0^2\lambda K_3}{\hbar c}.
\label{l_4}
\end{eqnarray}
\label{parameters}
\end{mathletters}
The London penetration depth $\lambda$ is given by
\begin{equation}
\frac{1}{\lambda^2}=\frac{32\pi e^2\eta_0^2}{\hbar^2 c^2}(K_1+K_2).
\end{equation}
In deriving Eq.~(\ref{Maxwell2}) we have replaced
\[
\frac{1}{\lambda^2}
\left[1-\frac{K_2}{K_1+K_2}\left(2e^{-x/\xi}-e^{-2x/\xi}\right)\right]
\]
by  $\lambda^{-2}$ ignoring the spatial dependence.
This approximation weakly affects a numerical factor in the final
expression of the Hall voltage.

For temperatures close to the onset of superconductivity,
$\tau=(T_c-T)/T_c$ is a small parameter which allows for a controlled
approximation.
From the standard temperature dependence of $\eta_0$,
$\xi$, and $\lambda$, it is clear that
\begin{equation}
\frac{1}{l_i}={\cal O}(\tau).
\label{order estimate}
\end{equation}

We first consider the case of vanishing external fields, i.e., 
$\bbox{E}=\bbox{B}=0$ in the vacuum ($x<0$).
This means that the net charge and current in the superconductor are 
zero.
In this case the scalar potential $A_0^{(0)}$ and the vector potential 
$A_y^{(0)}$ obey the boundary conditions
$dA_0^{(0)}/dx=dA_y^{(0)}/dx=0$ at $x=0$.
From Eq.~(\ref{order estimate}) we see that
$A_0^{(0)}={\cal O}(\tau^2)$ and $A_y^{(0)}={\cal O}(\tau^{1/2})$
since $\ell={\cal O}(\tau^0)$ and $\ell\ll\xi,\lambda$.
In lowest order in $\tau$ we find
\begin{eqnarray}
A_y^{(0)}(x)&=&
\varepsilon\frac{e}{l_4}
\frac{\kappa}{\kappa^2-1}
\left(\kappa e^{-x/\lambda}-e^{-x/\xi}\right),
\label{A_y^0}\\
A_0^{(0)}(x)&=&
-\ell^2\left\{
\frac{\varepsilon}{l_1}\left(1-e^{-x/\xi}\right)
\frac{dA_y^{(0)}(x)}{dx}
+\frac{\varepsilon}{\xi}\left(\frac{1}{l_1}+\frac{1}{l_2}\right)
 \left[A_y^{(0)}(x)e^{-x/\xi}
       -\frac{\ell}{\xi}A_y^{(0)}(0)e^{-x/\ell}\right]
-\frac{1}{e l_3(x)}\left[A_y^{(0)}(x)\right]^2
\right.\nonumber\\
&&\qquad\left.
+\frac{4\pi\eta_0^2}{\xi^2}
 \left[
   2C_1e^{-x/\xi}-\left(4C_1+\widetilde{K}_1\right)e^{-2x/\xi}
   +\frac{2\ell}{\xi}\left(3C_1+\widetilde{K}_1\right)e^{-x/\ell}
 \right]
\right\},
\label{A_0^0}
\end{eqnarray}
where $ \kappa = \lambda/\xi $.
This gives the total magnetic flux per length,
\begin{equation}
\Phi^{(0)}=\int^\infty_0dxB_z(x)=-A_y^{(0)}(0)
=-\varepsilon\frac{e}{l_4}
 \frac{\kappa}{\kappa+1}.
\end{equation}
Note that the vector potential $A_y^{(0)}$ is proportional to
$\varepsilon$ whereas the scalar potential is independent of the
chirality. For both the scalar and the vector potential the London
penetration depth constitutes the longest length scale of variation,
because they are coupled together.
The charge density $-(1/4\pi)d^2A_0^{(0)}/dx^2$ has
dipolar form as found in the BdG calculation. 
The total current density is
\begin{equation}
j_y^{(0)}=-\frac{c}{4\pi}\frac{d^2A_y^{(0)}(x)}{dx^2}
=\varepsilon\frac{ce}{4\pi l_4\lambda\xi}\frac{\kappa}{\kappa^2-1}
\left(\kappa e^{-x/\xi}-e^{-x/\lambda}\right),
\label{j_y^0}
\end{equation}
where the first term is related to the spontaneous current due to the
chiral  Andreev bound states, and the second term is the response of
the superconductor, i.e., the screening current.
Thus the net current is zero in the absence of an external magnetic
field.

Results obtained for the semi-infinite geometry can be easily
carried over to a Hall bar that extends from $x=0$ to $x=L_x$.
When the width $L_x$ of the Hall bar is much larger than $\lambda$,
the two surfaces are basically disconnected electromagnetically.
The gauge fields for the Hall bar geometry are then
obtained by combining the contributions from the two edges.
The scalar potential is $A_0^{(0)}(x)+A_0^{(0)}(L_x-x)$, while the
vector potential is $A_y^{(0)}(x)-A_y^{(0)}(L_x-x)$.

\subsection{Response to external fields}

We now consider two cases where a weak external field is applied to
the semi-infinite system. 
Weak perturbations introduce corrections to the gauge fields, 
$A_0=A_0^{(0)}+\delta A_0$ and $A_y=A_y^{(0)}+\delta A_y$.
We would like to obtain $\delta A_\mu$ in linear response to the
external perturbation. We can, therefore, linearize the Maxwell
equations in $ \delta A_{\mu} $:
\begin{eqnarray}
&&
\frac{d^2\delta A_0}{dx^2}-\frac{\delta A_0}{\ell^2}
-\frac{\varepsilon}{l_1}\left(1-e^{-x/\xi}\right)
 \frac{d\delta A_y}{dx}
-\frac{\varepsilon}{\xi}\left(\frac{1}{l_1}+\frac{1}{l_2}\right)
 e^{-x/\xi}\delta A_y
+\frac{2A_y^{(0)}}{e l_3(x)}\delta A_y=0,
\label{delta-Maxwell1}\\
&&
\frac{d^2\delta A_y}{dx^2}-\frac{\delta A_y}{\lambda^2}
-\frac{\varepsilon}{l_1}\left(1-e^{-x/\xi}\right)
 \frac{d\delta A_0}{dx}
+\frac{\varepsilon}{\xi l_2}e^{-x/\xi}\delta A_0
-\frac{2}{e l_3(x)}\left(A_0^{(0)}\delta A_y+A_y^{(0)}\delta A_0\right)
=0.
\label{delta-Maxwell2}
\end{eqnarray}

(1) In the first case a weak external magnetic field
$B\hat{\bbox{z}}$ is present in the vacuum, which corresponds to 
a finite net current running in the $y$ direction.
The boundary conditions are
$d\delta A_0/dx=0$ and $d\delta A_y/dx=B$ at $x=0$. 
In leading orders in $ \tau $ we obtain
\begin{eqnarray}
\delta A_y^{(1)}(x)&=&-\lambda Be^{-x/\lambda},
\label{A_y^1}\\
\delta A_0^{(1)}(x)&=&
-\varepsilon B\ell^2\left\{
\frac{1}{l_1}
\left[
 e^{-x/\lambda}\left(1-e^{-x/\xi}\right)
 +\frac{\ell}{\xi}e^{-x/\ell}
\right]
\right.\nonumber\\
&&\qquad\quad
-\kappa\left(\frac{1}{l_1}+\frac{1}{l_2}\right)
\left(
 e^{-x/\lambda-x/\xi}-\frac{\ell}{\xi}e^{-x/\lambda-x/\ell}
 -\frac{\ell}{\lambda}e^{-x/\ell}
\right)
\nonumber\\
&&\left.\qquad\quad
+\frac{2}{l_4} \frac{\kappa}{\kappa^2-1}
\left[
 \frac{\lambda}{l_3(x)}e^{-x/\lambda}
 \left(\kappa e^{-x/\lambda}-e^{-x/\xi}\right)
 -\frac{\ell}{l_3(0)}e^{-x/\ell}(\kappa-1)
\right]
\right\}.
\label{A_0^1}
\end{eqnarray}
We note that the sign of the scalar potential in linear response to
$B$ depends on the chirality $\varepsilon$. 

This result can be used to determine the transverse voltage in the 
Hall bar.
We assume that the induced current flows symmetrically on the two
edges.
Formally this situation is realized by applying the field
$B\hat{\bbox{z}}[\Theta(-x)-\Theta(x-L_x)]$.
The change of the scalar potential
is $\delta A_0^{(1)}(x)-\delta A_0^{(1)}(L_x-x)$, 
leading to a finite transverse voltage across the Hall bar,
$V_H=A_0(L_x)-A_0(0)=-2\delta A_0^{(1)}(0)$. 
Now we can make connection to the SHE in a two-dimensional system or
in a layered system like Sr$_2$RuO$_4$, where the system consists of
layers ($xy$-planes) separated by a distance $d$.
The total current per layer $I$ is related to the
external magnetic field by $I = cdB/2\pi$.
From Eq.~(\ref{A_0^1}) we find
\begin{equation}
V_H = \varepsilon I
\frac{4\pi\ell^2}{cd}
\left[-\kappa\left(\frac{1}{l_1}+\frac{1}{l_2}\right)
+\frac{2\lambda}{l_3(0)l_4}\frac{\kappa}{\kappa+1}\right],
\label{V-I-Hall}
\end{equation}
where we have kept only the leading terms, which are proportional to
$\tau$. The parameters entering this expression will be derived from a 
microscopic model in the weak-coupling limit in the next section. 

(2) We now turn to the second case where a weak external
electric field $E\hat{\bbox{x}}$ is applied to the superconductor,
inducing a finite surface charge.
A similar problem was recently discussed by Goryo and
Ishikawa.\cite{Goryo}
The boundary conditions here are 
$d\delta A_0/dx=-E$ and $d\delta A_y/dx=0$ at $x=0$. 
In lowest orders in $\tau$ we find the solution to
Eqs.~(\ref{delta-Maxwell1}) and (\ref{delta-Maxwell2}),
\begin{eqnarray}
\delta A_0^{(2)}(x)&=&E\ell e^{-x/\ell},
\label{A_0^2}\\
\delta A_y^{(2)}(x)&\approx&
-\varepsilon E\ell^2 e^{-x/\ell}
\left(
\frac{x+2\ell}{l_1\xi}+\frac{\ell}{l_2\xi}
-\frac{2\ell}{l_3(0)l_4}
 \frac{\kappa}{\kappa+1}
\right)
\nonumber\\&&
+\varepsilon E\ell^2e^{-x/\lambda}
\left[\kappa\left(\frac{1}{l_1}+\frac{1}{l_2}\right)
-\frac{2\lambda}{l_3(0)l_4}
 \frac{\kappa}{\kappa+1}
\right].
\label{A_y^2}
\end{eqnarray}
The external electric field changes the local configuration of
electric current and magnetic field in the chiral superconductor.
The induced vector potential $\delta A_y^{(2)}$ depends on the
chirality $\varepsilon$.
The total change in the magnetic flux (per unit length along the $y$ 
direction at the surface) is
\begin{equation}
\delta \Phi = - \delta A_y^{(2)}(0) =
\varepsilon E\ell^2
\left[
-\kappa\left(\frac{1}{l_1}+\frac{1}{l_2}\right)
+\frac{2\lambda}{l_3(0)l_4}
 \frac{\kappa}{\kappa+1}
\right].
\label{flux change}
\end{equation}
Note the similarity between Eqs.~(\ref{V-I-Hall}) 
and (\ref{flux change}) indicating that the two phenomena have indeed
common origin.

\section{The weak-coupling coefficients of the GL theory}

In this section we calculate the coefficients of the GL free energy in
the weak-coupling limit using a model with layers of two-dimensional
electron gas, where electrons are confined in each layer.
Furthermore we ignore any spatial variation of electromagnetic fields
in the direction perpendicular to the layers.

The mean-field Hamiltonian appropriate for the discussion of the
chiral $p$-wave state (\ref{chirp}) is already given in
Eq.~(\ref{MFHamiltonian}). 
The static electromagnetic fields are governed by
\begin{equation}
{\cal H}_{\rm EM}
=\frac{d}{8\pi}
 \left[(\nabla\times\bbox{A})^2 - (\nabla A_0)^2\right],
\label{H_EM}
\end{equation}
where $d$ is the distance between the two-dimensional layers.
The sign of the second term was chosen negative so that the Maxwell
equations can be obtained by taking the functional
derivative of
$\int d^2r\left({\cal H}_{\rm MF}+{\cal H}_{\rm EM}\right)$
with respect to $A_\mu$, 
which should be viewed as the Lagrangian in imaginary time.

We integrate out the electron fields $\psi_\sigma$ and
$\bar\psi_\sigma$ to obtain the effective functional $F_{\rm eff}$
for $\bbox{\eta}$ and $A_\mu$:
\begin{equation}
\exp(-F_{\rm eff}/k_BT)
=\int\prod_\sigma{\cal D}\bar\psi_\sigma{\cal D}\psi_\sigma
\exp\left[
-\frac{1}{\hbar}\int^{\hbar/k_BT}_0dt\int d^2r
 \left(\hbar \bar\psi_\sigma\partial_t\psi_\sigma
        + {\cal H}_{\rm  MF} + {\cal H}_{\rm EM} \right)
\right],
\label{Z}
\end{equation}
where the electron field operators $\psi_\sigma(\bbox{r})$
and $\psi_\sigma^\dagger(\bbox{r})$
in ${\cal H}_{\rm MF}$ (\ref{MFHamiltonian}) are
replaced by Grassman fields $\psi_\sigma(\bbox{r},it)$
and $\bar\psi_\sigma(\bbox{r},it)$, respectively.
The GL equations are then obtained by taking the
functional derivative of $F_{\rm eff}$ with respect to
$\bbox{\eta}$ and $A_\mu$.
We calculate $F_{\rm eff}$ in powers of $\eta_i$ and $A_\mu$ up to the
order $\eta^4_i$, $D_i\eta_jD_k\eta_l$, $ A_0 D_i\eta_jD_k\eta_l$ and
$E_i\eta_jD_k\eta_l$ in the 
weak-coupling limit, $|\bbox{\eta}|\ll\hbar^2k^2_F/2m$.
The calculation is tedious but straightforward, and only the
final result is presented below.
The functional $F_{\rm eff}$ so obtained has the form
\begin{equation}
F_{\rm eff}=
\int d^3r\left({\cal F}+{\cal{L}}_e \right).
\label{F_eff}
\end{equation}
The free energy part $ {\cal F} $, which is defined in
Eq.~(\ref{standardfe}) as the free-energy density in {\em three}
dimensions, has the standard coefficients,
\begin{mathletters}
\begin{eqnarray}
&&
a  = -\frac{\tau}{2}N(0), \\
&&
\frac{b_1}{3}= \frac{b_2}{2} = - \frac{b_3}{2}
= \frac{7 \zeta(3) N(0)}{128(\pi k_B T_c)^2}, \\
&&
\frac{K_1}{3} = K_2 = K_3 = K_4
= \frac{7 \zeta(3) N(0)}{128}
  \left(\frac{\hbar v_F}{\pi k_B T_c} \right)^2,
\label{Ks}
\end{eqnarray}
\label{coeff1}
\end{mathletters}\noindent
where $ T_c $ depends on the coupling constant $ g $ in the usual
exponential form
$ k_B T_c = \omega_c \exp[- 2/ g N(0)]$ with $\omega_c$ being the
cutoff energy scale, and
$ N(0) = m/(2\pi\hbar^2d)$ is the density of states (per spin) at the
Fermi level.
The results (\ref{coeff1}) are valid in lowest order in
$|\bbox{\eta}|/\mu$, where we find $K_3=K_4$.
If we assume that the density of states had a weak energy dependence
with energy derivative
 $N'(0)\approx N(0)/\mu=m^2/(\pi\hbar^4 k_F^2d)$
at the Fermi surface, then there would be a tiny difference between
$K_3$ and $K_4$, yielding the contribution to the free energy density,
\begin{equation}
{\cal F}_{\rm Z}=
\frac{e\hbar N'(0)}{8cm}
 \ln\left(\frac{2e^{\bf C}\omega_D}{\pi T_c}\right)
 [i\bbox{\eta}(\bbox{r})\times\bbox{\eta}^*(\bbox{r})]\cdot\bbox{B}
=
-\frac{n_0\eta_0^2}{2\mu^2}
\ln\left(\frac{2e^{\bf C}\omega_D}{\pi T_c}\right)
\bbox{\mu}\cdot\bbox{B},
\label{Zeeman}
\end{equation}
where ${\bf C}=0.5772\ldots$ is the Euler's constant,
$n_0=k_F^2/2\pi d$ is the electron density,
and $\bbox{\mu}$ is the magnetic moment (per electron) of a Cooper
pair in the chiral $p$-wave state,
\begin{equation}
\bbox{\mu}=
-\hat{\bbox{z}}\frac{\mu_B}{2\eta_0^2}{\rm Im}(\eta_x^*\eta_y).
\label{intrinsic moment}
\end{equation}
Here $\mu_B=e\hbar/2mc$ is the Bohr magneton.
Equation (\ref{Zeeman}) is the Zeeman energy for the intrinsic
magnetic moment of the chiral $p$-wave state.
As is well known,\cite{CROSS,MINEEV,MERMIN} it is diminished 
by the factor $(\eta_0/\mu)^2$ which indicates the degree of  
particle-hole asymmetry at the Fermi level. Since this contribution
is very small, we can ignore it in the following analysis.

From the coefficients in Eq.~(\ref{coeff1}) we immediately obtain
the order parameter,
\begin{equation}
\eta_0(T)
=\pi k_B T_c\sqrt{\frac{8\tau}{7 \zeta(3)}},
\end{equation}
the coherence length,
\begin{equation}
\xi(T)
=\sqrt{\frac{2K_1}{|a|}}
=
\frac{\hbar v_F}{\pi k_B T_c}
\sqrt{\frac{21 \zeta(3)}{32\tau}},
\end{equation}
and the London penetration depth,
\begin{equation}
\lambda(T)=
\frac{\hbar c}{2e\eta_0}
\sqrt{\frac{1}{8\pi(K_1+K_2)}}
=\frac{c}{ev_F}\sqrt{\frac{1}{8\pi\tau N(0)}}.
\label{lon-pen}
\end{equation}
The second contribution to the free energy, ${\cal L}_e$, has the
following coefficients:
\begin{mathletters}
\begin{eqnarray}
&&
\frac{1}{8\pi\ell^2}=e^2N(0),
\label{Thomas-Fermi}\\
&&
\frac{\widetilde{K}_1}{3} = \widetilde{K}_2 = \widetilde{K}_3 =
\widetilde{K}_4 =
\frac{7 \zeta(3) e}{128 \pi d (\pi k_B T_c)^2},
\label{tildeK} \\
&&
-C_1 = C_2 = -\frac{C_3}{3} = C_4 =
\frac{7\zeta(3)e}{256\pi d(\pi k_B T_c)^2}.
\label{Cs}
\end{eqnarray}
\label{coeff2}
\end{mathletters}
Notice that $\widetilde K_i=e\partial K_i/\partial\mu$.
This relation can be easily understood, if we regard
$eA_0$ as a spatial variation of $\mu$.
In the course of deriving Eqs.~(\ref{tildeK}) and (\ref{Cs}) 
we have naturally assumed that the momentum $\bbox{k}$ in
$\bbox{d}(\bbox{k})=\hat{\bbox{z}}(\bbox{\eta}\cdot\bbox{k})/k_F$ is
close to the Fermi surface.
The CS-like term describing the reaction of the intrinsic magnetic
moment to a change in the superfluid density does {\em not} have the
reduction factor found in Eq.~(\ref{Zeeman}), in agreement with the
argument by Volovik and Mineev.\cite{MINEEV,Volovik3}

From Eqs.~(\ref{coeff1}) and (\ref{coeff2}) we obtain the parameters
appearing in Eqs.~(\ref{V-I-Hall}) and (\ref{flux change}):
\begin{mathletters}
\begin{eqnarray}
&&
-l_1=\frac{l_2}{2}=\frac{\hbar cd}{2e^2\tau},
\\
&&
\frac{\lambda}{l_3(0)l_4}
=\frac{3e^2\tau}{4\hbar cd}.
\end{eqnarray}
\end{mathletters}

Now we may express the results for the two cases discussed in the last
section in terms of the microscopic parameters. 
The Hall voltage induced by an external current is found to be
\begin{equation}
V_H = \varepsilon I
 \frac{h}{16 (ek_F \lambda)^2}
 \left(2\kappa+\frac{3\kappa}{\kappa+1}\right)
\label{gl-trans-volt}
\end{equation}
in the leading order in $ \tau $.
Here we have made use of the relation
$(k_F\lambda)^2=mc^2d/4e^2\tau$.
We find a strong reduction compared with the quantum unit of
resistance $ R_0 = h/2e^2 $.
The factor $1/(k_F\lambda)^2$ can be also written as
$16 \pi n_s\chi/n_0$, where $n_s$ is the superfluid density and
$\chi=\mu_B^2N(0) = 3 \chi_{\rm orb}$ corresponding to the orbital
susceptibility. 
We can thus rewrite Eq.~(\ref{gl-trans-volt}) as
\begin{equation}
V_H=\varepsilon I\frac{ \pi h\gamma\chi}{e^2}
\left(2\kappa+\frac{3\kappa}{\kappa+1}\right),
\label{V_H-GL}
\end{equation}
where $\gamma=n_s/n_0=2\tau$ is the ratio of the superfluid density to
the electron density. Obviously the Hall resistance strongly depends on 
material dependent parameters and is also rather severely reduced from
the universal value $ h / e^2 $.
In Sec.~VI we will analyze the quantitative aspect in more detail. 

We turn now to the reciprocal case (2).
Here the effect is more subtle as the response to an external electric
field constitutes a change of the field distribution in the vicinity
of the surface.
Since there is already spontaneous magnetization generated by the
surface currents, we estimate the relative change of the total
magnetic flux:
\begin{equation}
\frac{\delta \Phi}{\Phi^{(0)}} =
\frac{\delta A_y^{(2)}(0)}{A_y^{(0)}(0)}
=-\frac{eE\ell^2}{\mu\lambda}\left(\kappa+\frac{5}{2}\right).
\end{equation}
The electric field $E$ should be smaller than
$\Delta/e\ell$ in order to avoid nonlinear effects that arise from the
field effect on the superconducting order parameter at the
surface. Therefore the induced change of flux $\delta\Phi$ would 
only be a very small fraction of the spontaneous magnetic flux
$\Phi^{(0)}$.  

\section{Discussion based on hydrodynamic equations}

\subsection{The hydrodynamic equations} 

In the previous sections we have considered the SHE
using the self-consistent solution of the BdG equation at $T = 0$ and
the extended GL theory near $T_c$.
In this section we introduce a phenomenological description based on
the stationary hydrodynamic equations.
This approach can provide an interpolation between the two limits
and allow us to have a simple intuitive understanding of the physics
involved. 
Our starting point is the phenomenological Lagrangian
\begin{equation}
F=\int d^3r\left[
f+eA_0\frac{\partial f}{\partial \mu}
-\frac{1}{8\pi}\left(\nabla A_0\right)^2
-\frac{1}{8\pi\ell^2}A_0^2
+\frac{1}{8\pi}\left(\nabla\times\bbox{A}\right)^2
\right],
\label{F}
\end{equation}
which describes electromagnetic properties of a superconductor.
In the chiral $p$-wave state the Lagrangian density $f$ may be
written as
\begin{equation}
f=n_s\left[
\frac{m}{2}(\bbox{v}_s+\bbox{v}_0)^2
+\nabla\cdot(\bbox{\mu}\times\bbox{A})
\right],
\label{f}
\end{equation}
where $\bbox{v}_s=(e/mc)\bbox{A}$ and $\bbox{v}_0$ is the velocity of
the supercurrent generated by the spatial dependence of the order
parameter:
\begin{equation}
\bbox{v}_0= \left( \begin{array}{c} v_{0x} \\ v_{0y} \end{array}
\right) = 
\frac{\hbar}{8m\eta_0^2} \left[ \pmatrix{ {\rm Im} \left(3 \eta_x^*
\partial_x \eta_x + \eta_y^* \partial_x \eta_y \right)
\cr
{\rm Im} \left(3 \eta_y^*
\partial_y \eta_y + \eta_x^* \partial_y \eta_x \right)
\cr} + 
\pmatrix{
{\rm Im}\left(\eta_x^*\partial_y\eta_y+\eta_y^*\partial_y\eta_x\right)
\cr
{\rm Im}\left(\eta_x^*\partial_x\eta_y+\eta_y^*\partial_x\eta_x\right)
\cr
} \right].
\label{spont-velo}
\end{equation}
The first term corresponds to the ordinary supercurrent due to phase
gradient, while the second term is connected
with the spontaneous current due to texture of the order parameter. 
The latter is equivalent to the surface current of the chiral edge
states and will be important for our discussion.
The partial derivative in Eq.~(\ref{F}) acts on $n_s$ as
\begin{equation}
\frac{\partial n_s}{\partial \mu}=
\frac{\partial n_s}{\partial n_0}\frac{\partial n_0}{\partial\mu}
=2\gamma N(0),
\end{equation}
where $\gamma= n_s/ n_0$ is the ratio of the
superfluid density $n_s$ to the electron density $n_0=k_F^2/2\pi d$.
We know the two limiting values of $\gamma$: $\gamma=1$ at $T=0$ and
$\gamma=2\tau$ near $T_c$.

The Lagrangian density $f$ can be deduced from the GL Lagrangian in
the following way.
In the chiral $p$-wave state we may assume without loss of generality
that $\eta_x^* \eta_y $ is imaginary.
With the weak-coupling result (\ref{Ks}),
it is easy to confirm that the $K_i$ terms in ${\cal F}$
(\ref{standardfe}) generate $n_sm\bbox{v}_s\cdot\bbox{v}_0$,
in addition to $n_s(m/2)\bbox{v}_s^2$.
The latter term can be obtained with the approximation
$|\eta_x|^2 =|\eta_y|^2 =\eta_0^2$.
It is then natural to complete the square to make
$m(\bbox{v}_s+\bbox{v}_0)^2/2$, which is nothing but the kinetic
energy of the superfluid in the presence of the spontaneous flow with
the velocity $\bbox{v}_0$.
From the relation $\widetilde K_i=e\partial K_i/\partial\mu$, one can
also see that the $\widetilde K_i$ terms lead to the kinetic energy
contribution to $eA_0\partial f/\partial\mu$.
Furthermore, we find from Eqs.~(\ref{edfe}) and (\ref{Cs}) that
the $C_i$ terms have a contribution
\begin{eqnarray}
&&
-iC_1\frac{8e}{\hbar c}\int d^3r
(\eta^*_x\eta_y-\eta^*_y\eta_x)(A_y\partial_xA_0-A_x\partial_yA_0)
\nonumber\\&&\qquad
=iC_1\frac{8e}{\hbar c}\int d^3r
A_0\nabla\cdot\left[
 \bbox{A}\times\hat{\bbox{z}}(\eta_x^*\eta_y-\eta_y^*\eta_x)
\right]
+{\rm surface~term}
\end{eqnarray}
in the chiral $p$-wave state where $\eta_x^* \eta_y $ is 
imaginary.
This gives the remaining term,
$2\gamma eA_0N(0)n_s\nabla\cdot(\bbox{\mu}\times\bbox{A})$.
With the help of the identity
$\nabla\cdot(\bbox{\mu}\times\bbox{A})=
 \bbox{A}\cdot(\nabla\times\bbox{\mu})
-\bbox{\mu}\cdot(\nabla\times\bbox{A})$,
we find that $\nabla\cdot(\bbox{\mu}\times\bbox{A})$
represents the coupling of the magnetic current
$\nabla\times\bbox{\mu}$ to the vector potential
as well as the Zeeman energy of the magnetic moment $\bbox{\mu}$.
It is important to realize that the full magnetic moment of a chiral 
Cooper pair appears here, without reduction found in
Eq.~(\ref{Zeeman}).
The scalar potential induces a change in the number of Cooper pairs,
which is necessarily accompanied by the change of the full magnetic
moment per Cooper pair.\cite{MINEEV}

We now take the variation of $F$ with respect to $\bbox{A}$ to obtain
the extended London equation,
\begin{equation}
\nabla^2\bbox{A}-\frac{1}{\lambda^2}\bbox{A}
 +\frac{4\pi}{c}\bbox{j}=0,
\end{equation}
with the current density
\begin{equation}
\bbox{j}=-n_s e\bbox{v}_0
-2e\gamma N(0)[eA_0(\bbox{v}_s+\bbox{v}_0)+c\bbox{E}\times\bbox{\mu}].
\end{equation}
The variation of $F$ with respect to $A_0$ yields
\begin{equation}
\nabla^2A_0-\frac{1}{\ell^2}A_0+\frac{8\pi eN(0)}{n_0}f=0.
\label{variation-A_0}
\end{equation}
Since the Thomas-Fermi screening length $\ell$ is much shorter than
$\xi$ and $\lambda$, we may ignore $\nabla^2A_0$ to obtain immediately 
\begin{equation}
A_0=
\frac{n_s}{en_0}\left[
\frac{m}{2}\left(\bbox{v}_s+\bbox{v}_0\right)^2
+\nabla\cdot(\bbox{\mu}\times\bbox{A})
\right],
\label{Bernoulli}
\end{equation}
which we may call a generalized Bernoulli equation.
The first term represents a Bernoulli force coming from the kinetic
energy of superfluid.
In conventional superconductors without broken time-reversal symmetry,
the spontaneous current is absent.
In such a case a supercurrent $I$ injected from an external current
source can generate a transverse potential difference proportional to
$I^2$.\cite{LONDON,STAAS,RICK}
On the other hand, in the case of our interest where a spontaneous
current flows along a boundary ($\bbox{v}_0\ne0$), an external current
can induce a transverse potential difference proportional to $I$.
The second term in (\ref{Bernoulli}) is characteristic of the chiral
$p$-wave state where Cooper pairs have their magnetic moment.
Both terms are important in the SHE and in its reciprocal effect.

\subsection{Spontaneous Hall effect and its reciprocal effect}

Let us study the spontaneous Hall effect using the hydrodynamic
equations.
As in Sec.~IV, we consider a chiral $p$-wave superconductor with
a boundary at $x=0$ only.
The superconductor occupies the positive $x$ region and the system is
translationally invariant in the $y$ and $z$ directions.
We suppose that the spatial profile of the order parameter
$\bbox{\eta}(\bbox{r})=\biglb(\eta_x(x),\eta_y(x)\bigrb)$ is already
determined self-consistently in the equilibrium state.
In particular, as we have observed in the BdG calculation in Sec.~II,
the order parameter satisfies
$\eta_x=d\eta_y/dx=0$ at $x=0$ under the assumption of
the specular reflection at the surface.
We will calculate a linear response to week electromagnetic 
perturbation.

The local variation of $\bbox{\eta}$ near the boundary yields a
spontaneous current due to the order parameter texture,
$\bbox{j}_t=-n_se\bbox{v}_t$, where 
\begin{equation}
\bbox{v}_t
=\frac{\hbar}{8m\eta_0^2}
 {\rm Im}\left(
  \eta_x^*\frac{d\eta_y}{dx}+\eta_y^*\frac{d\eta_x}{dx}
 \right)\hat{\bbox{y}}.
\label{v_t}
\end{equation}
This current determines the vector potential $\bbox{A}=(0,A_y,0)$
obeying the London equation,
\begin{equation}
\left(\frac{d^2}{dx^2}-\frac{1}{\lambda^2}\right)\bbox{A}
=-\frac{4\pi}{c}\bbox{j}_t,
\end{equation}
whose solution under the boundary condition
$\partial_x A_y=0$ at $x=0$ is
\begin{equation}
\bbox{A}^{(0)}(x)=\frac{2\pi\lambda}{c}\int^\infty_0
\left(e^{-|x-x'|/\lambda}+e^{-|x+x'|/\lambda}\right)
\bbox{j}_t(x')dx'.
\label{A-solution}
\end{equation}
In particular, its boundary value is
\begin{equation}
A_y^{(0)}(0)=
\frac{\pi n_s\lambda\mu_B}{\eta_0^2}
\int^\infty_0e^{-x/\lambda}{\rm Im}
\left(\eta_x\frac{d\eta_y^*}{dx}+\eta_y\frac{d\eta_x^*}{dx}\right)dx.
\label{A_y^0(0)}
\end{equation}
Having determined $\bbox{\eta}$ and $\bbox{A}$, we are now ready to
obtain the scalar potential $A_0$ from the generalized Bernoulli
equation.

Suppose that a small current $I$ is injected from an external source
to the superconductor, yielding a small change in the vector
potential: $\bbox{A}=\bbox{A}^{(0)}+\delta\bbox{A}$, where
$\delta\bbox{A}(0)=-I(2\pi\lambda/cd)\hat{\bbox{y}}$.
Accordingly, the scalar potential acquires a small change,
whose boundary value is
\begin{equation}
\delta A_0(0)=\left.
\frac{n_s}{en_0}\delta\bbox{A}\cdot\left(
\frac{e^2}{mc^2}\bbox{A}^{(0)}+\frac{e}{c}\bbox{v}_t
+\nabla\times\bbox{\mu}\right)
\right|_{x=0},
\label{deltaA_0}
\end{equation}
where we have used the fact that $\bbox{\mu}=0$ at $x=0$.
The intrinsic magnetic moment $\bbox{\mu}$ defined in
Eq.~(\ref{intrinsic moment}) gives
\begin{equation}
\nabla\times\bbox{\mu}=
\frac{\mu_B}{2\eta_0^2}{\rm Im}\left(
\frac{d\eta_x^*}{dx}\eta_y+\eta_x^*\frac{d\eta_y}{dx}
\right)\hat{\bbox{y}}.
\label{dmu/dx}
\end{equation}
The spontaneous Hall voltage in the Hall bar geometry can be related
to $\delta A_0$ by the relation $V_H=-2\delta A_0(0)$ as in Sec.~IV.
From Eqs.~(\ref{v_t}), (\ref{A-solution}), and (\ref{dmu/dx}) we
obtain
\begin{equation}
\frac{V_H}{I}=\frac{2\pi h n_s\chi}{e^2 n_0\eta_0^2}
\left[
\left.\lambda{\rm Im}
     \left(\eta_y\frac{\eta_x^*}{dx}\right)\right|_{x=0} 
+\int^\infty_0 e^{-x/\lambda}
 {\rm Im}\left(\eta_x\frac{\eta_y^*}{dx}
              +\eta_y\frac{\eta_x^*}{dx}\right)dx
\right].
\label{V_H-from-hydro}
\end{equation}
With the approximate form (\ref{approximate-eta_x}) of the order
parameter used in Sec.~IV, Eq.~(\ref{V_H-from-hydro}) becomes
\begin{equation}
\frac{V_H}{I}=
\varepsilon\gamma\chi\frac{2\pi h}{e^2}
\left(\kappa+\frac{\kappa}{\kappa+1}\right)
\label{V_H-from-hydro-2}
\end{equation}
which compares well with (\ref{V_H-GL}).

The quantitative comparison with the self-consistent solution of the
BdG equations encounters some drawback from the fact that for the sake
of numerical accuracy the cutoff energy $\omega_c$ had to be taken
comparable to the Fermi energy.
Therefore, there are larger strong-coupling corrections not included
in our phenomenological analysis.
Nevertheless, we find a reasonably good agreement between the
phenomenological and the BdG estimates.
The result from the numerical 
BdG calculation ($ k_F \xi_0 = 16 $ and $ \ell/\xi_0 = 0.1 $) is
\begin{equation}
\frac{V_H}{I} \approx \frac{h}{2 e^2} \times 10^{-3} \times
\left\{ \begin{array}{ll} 1.3  \qquad \qquad & \kappa =1, \\
0.5 & \kappa =2, \end{array} \right.
\end{equation}
while Eq.~(\ref{V_H-from-hydro-2}) with the same parameters (apart
from $\omega_c$) leads to
\begin{equation}
\frac{V_H}{I} \approx \frac{h}{2 e^2} \times 10^{-3} \times
\left\{ \begin{array}{ll} 1.5 \qquad \qquad & \kappa =1, \\
0.65 & \kappa =2. \end{array} \right.
\end{equation}
The discrepancy is not only a result of weak-coupling versus
strong-coupling approach, but we would like to remind that we have
also used an approximate description of the order parameter texture
at the surface. 

Obviously the resistance obtained here is considerably smaller than
the universal unit $ h/e^2 $.
We would like to build now a connection between the SHE and the
ordinary Hall effect.
We may look for the intrinsic magnetic field which causes the Hall
response to the externally induced current $I$.
Comparing Eq.~(\ref{V_H-from-hydro-2}) 
with the standard expression of the Hall effect, 
\begin{equation}
V_H = \frac{1}{n_0 e c} \frac{I H_{\rm eff}}{d},
\end{equation}
we obtain the effective magnetic field
\begin{equation}
H_{\rm eff} = \pi n_s \mu_B \kappa  \left( 1 + \frac{1}{1 + \kappa}
\right).
\end{equation}
The effective field corresponds to the density of
magnetic moments of Cooper pairs which is not the reduced magnetic
moment in (\ref{Zeeman}), but rather the full moment which is
associated with the change of the Cooper pair density.  
Note that the factor $ \pi \kappa [1 + 1/(1+\kappa)] $ is due to
inhomogeneous field and current distribution.

Next we consider the reciprocal effect.
In the presence of a weak external electric field
$\delta\bbox{E}=E\hat{\bbox{x}}$ at $x<0$,
the scalar potential in the superconductor receives a small
perturbation, $\delta A_0(x)=E\ell e^{-x/\ell}$.
This yields a small change in the current density,
\begin{eqnarray}
\delta\bbox{j}(x)&=&
-2e\gamma N(0)\left[
 e\delta A_0\left(\bbox{v}_t+\frac{e}{mc}\bbox{A}^{(0)}\right)
 +c\delta\bbox{E}\times\bbox{\mu}
\right]
\nonumber\\
&=&
-2ce\gamma N(0)Ee^{-x/\ell}\left[
\ell\left(\frac{e}{c}v_{ty}+\frac{e^2}{mc^2}A^{(0)}_y\right)
-\mu_z\right]\hat{\bbox{y}},
\end{eqnarray}
which determines the change in the vector potential $\delta\bbox{A}$.
Solving the London equation, we find
\begin{equation}
\delta A_y(0)=\frac{4\pi\lambda}{c}\int^\infty_0e^{-x/\lambda}
\delta j_y(x)dx.
\end{equation}
Since the Thomas-Fermi screening length $\ell$ is much shorter than
$\xi$ and $\lambda$, we may set
$v_{ty}(x)=v_{ty}(0)$, $A^{(0)}_y(x)=A^{(0)}_y(0)$, and
$\mu_z(x)=x\mu'_z(0)$ in $\delta j_y(x)$ in the integrand.
With this approximation we obtain
\begin{eqnarray}
\delta A_y(0)&=&
-E\frac{\gamma\lambda}{e}
\left.
\left(
\frac{e}{c}v_{ty}+\frac{e^2}{mc^2}A_y^{(0)}-\frac{d\mu_z}{dx}
\right) 
\right|_{x=0}
\nonumber\\
&=&
-E\frac{n_s\mu_B}{4en_0\eta_0^2}
\left[
\left.
\lambda{\rm Im}\left(\eta_y\frac{\eta_x^*}{dx}\right)
\right|_{x=0}
+\int^\infty_0 e^{-x/\lambda}
 {\rm Im}\left(\eta_x\frac{\eta_y^*}{dx}
              +\eta_y\frac{\eta_x^*}{dx}\right)dx
\right].
\label{delta A_y(0)}
\end{eqnarray}
The magnitude of the ratio $\delta A_y(0)/A^{(0)}_y(0)$ may be
estimated from Eqs.~(\ref{A_y^0(0)}) and (\ref{delta A_y(0)}) for the
approximate form (\ref{approximate-eta_x}), yielding
\begin{equation}
\frac{\delta A_y(0)}{A^{(0)}_y(0)}
=-\frac{eE\ell^2}{\mu\lambda} (\kappa + 2),
\end{equation}
in good agreement with the GL analysis. Unfortunately, in this case
the comparison with the BdG result does not agree well, which we 
attribute to the fact that $ \ell $ is comparable to the other length
scales in the numerical calculation. Thus, there are strong correction
to the above result in addition to the strong-coupling corrections.

\subsection{Experimental probe}

Although the phenomenon we discuss in this paper can be compared to
the standard Hall effect, it has actually some quite distinct
aspects. Flowing currents are an equilibrium property of the
superconductor. It is, therefore, impossible to measure the
transverse voltage by means of a standard
voltmeters, using direct contacts to the surfaces.
This problem was realized already more than
30 years ago, when the Bernoulli response to a current was
investigated in conventional superconductors.\cite{MORRIS} 
In this case a potential difference between the surface and the
interior of the superconductor is expected, which is proportional to
the square of the running current:
\begin{equation}
V = \frac{2 \pi}{e c^2 d^2 n_0} I^2. 
\label{ordinarybern}
\end{equation}
The potential difference is actually independent of temperature as has
been observed\cite{MORRIS} for Pb.
The method of measurement was based on a thin-film capacitor
which picks up the voltage signal caused by an ac-current on the
superconductor surface.
The same kind of capacitor technique could also be used to detect the
spontaneous Hall effect.
In an ac-measurement the above Bernoulli force ($\propto I^2$)
yields the second harmonic of the applied ac-current, while the signal
corresponding to the SHE ($\propto I$) contributes
to the first harmonic.
Hence we can distinguish the SHE from the
standard Bernoulli effect.
We can estimate the magnitude of the Hall response 
\begin{equation}
V_H = R_H I = \frac{R_H I_{\rm tot}}{N_{\rm layer}}
 \approx  \frac{h}{4e^2} \frac{1}{k_F^2 \lambda^2}
          \frac{I_{\rm tot}}{N_{\rm layer}}, 
\end{equation}
where $I_{\rm tot}$ is the total current running through a
three-dimensional sample consisting of $N_{\rm layer}$ layers.
Using typical values for ordinary metals (for example,
$k_F\lambda\approx200$ at $T=0$),
we obtain $R_H \approx 0.16 \Omega$.
If $I_{\rm tot}/N_{\rm layer}=1\,$nA, then we expect
$V_H\approx 0.16$nV, which might be experimentally accessible.
Under the same conditions the conventional
Bernoulli signal is considerably smaller, $ V \sim 1\,$pV.
It is worth noting that the capacitor technique in measuring the
transverse potential change does not require a Hall bar geometry, but
a single surface is sufficient.

Unfortunately, besides the high sensitivity necessary in this kind of
measurements a further problem has to be taken into account.
This is the formation of domains of degenerate
superconducting states with opposite chiralities.
Such domain formation is very likely to occur when a system enters
the superconducting state.
The two domains with opposite chiralities yield opposite sign of the
transverse voltage so that the net effect might be diminished.
Since the spontaneous Hall voltage is a surface effect, the number of
domain walls intersecting the surface matters.
It would be necessary to establish an experimental technique to realize
a single domain phase, for example, by cooling in a weak magnetic
field.

\section{Non-chiral time-reversal symmetry breaking states}

We now consider the possibility of a spontaneous Hall effect in
non-chiral time reversal symmetry breaking superconductors.
While for these superconducting states there is no anomalous CS-like
coupling between scalar and vector potential, there are still
spontaneous surface currents for certain orientations of the sample
boundaries, despite the fact that the Cooper pairs do not have a net
angular momentum.
These surface currents can be associated with Andreev bound states.
Thus we would expect at least to find a contribution to
the SHE due to the Bernoulli force.
We consider one of the well-known
examples of a non-chiral superconducting state in a
quasi-two-dimensional system that violates time-reversal symmetry,
the $ d+is $-wave state.
We assume that the $d$-wave state has the $d_{xy}$-symmetry, for which 
a spontaneous surface current runs along surfaces normal to the
[100] or [010]-direction.\cite{d+is}
It is possible here to express
the spontaneous current again as a result of the order parameter
texture analogous to (\ref{spont-velo}): 
\begin{equation}
{\bf v}_0 = \left( \begin{array}{c} v_{tx} \\ v_{ty} \end{array}
\right) = \frac{\hbar}{8 m \eta_0^2} \left( \begin{array}{c} 
{\rm Im}(\eta_s^* \partial_y \eta_d + \eta_d^* \partial_y \eta_s) \\
{\rm Im}(\eta_s^* \partial_x \eta_d + \eta_d^* \partial_x \eta_s) 
\end{array} \right),
\end{equation}
where $ \eta_s $ and $ \eta_d $ denote the order parameters of the
$s$-wave and $ d $-wave component respectively. 
The currents can be of similar magnitude as for the chiral $ p$-wave
state and consequently the size of the SHE is comparable. 
There are, of course, some differences from the former case. 
Since the two pairing components are not degenerate in general,
various additional parameters may appear in the discussion. 
We have performed a BdG calculation for a specific set of parameters
to verify the expectation of the above argument. 

\begin{figure}[t]
\begin{center}
\begin{minipage}[t]{8cm}
\epsfxsize=8cm
\epsfbox{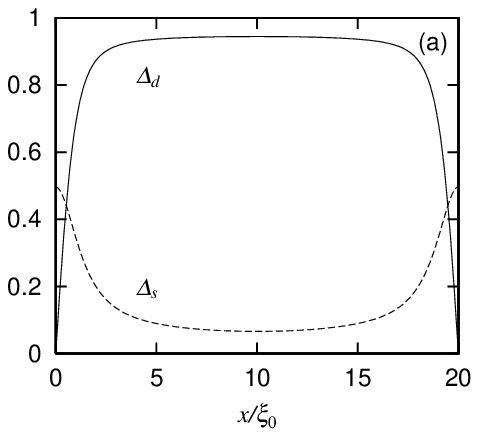}
\end{minipage}
\begin{minipage}[t]{8cm}
\epsfxsize=8cm
\epsfbox{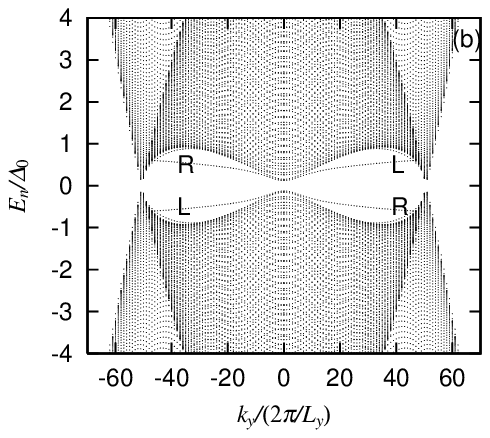}
\end{minipage}
\begin{minipage}[t]{8cm}
\epsfxsize=8cm
\epsfbox{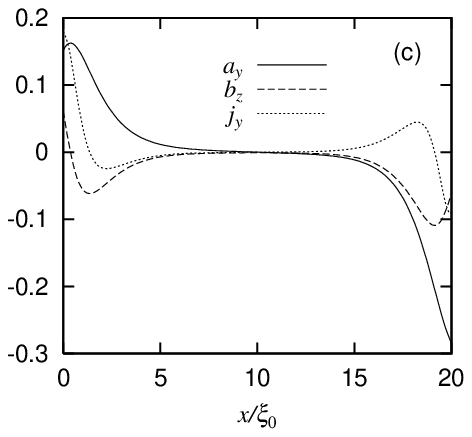}
\end{minipage}
\begin{minipage}[t]{8cm}
\epsfxsize=8cm
\epsfbox{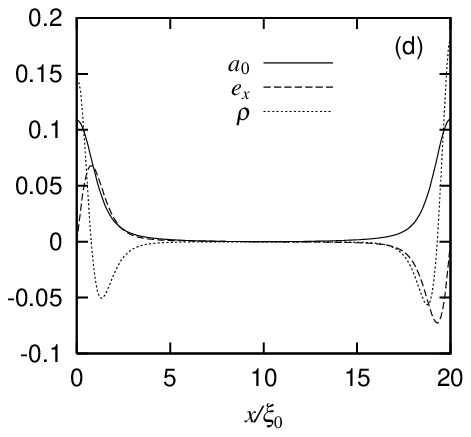}
\end{minipage}
\end{center}
\caption{Self-consistent solution of the Bogoliubov-de Gennes equation
for the [100] surface in the $d_{xy}+i s$-wave state at $T=0$.
The set of parameters are the same as used in Fig.~\ref{fig:BdG2}.
(a) Order parameter scaled by $\Delta_0$,
which is the magnitude of the order parameter for the
$d_{xy}$-wave in the bulk region. 
$\Delta_d$ and $\Delta_s$ are the magnitude of the order parameter
of the $d_{xy}$-wave and $s$-wave, respectively.
(b) Energy eigenvalues obtained for sine basis functions.
L and R denote the surface bound states localized near left and right
side surfaces, respectively.
(c) Dimensionless vector potential $a_y=(e\xi_0/\hbar c)A_y$,
magnetic field $b_z=\xi_0\partial_x a_y$
and current density $j_y=-\xi_0\partial_x b_z$.
Total current is $I=5e\Delta_0/\hbar k_F\xi_0$.
(d) Dimensionless scalar potential $a_0=eA_0/\Delta_0$,
electric field $e_x=-\xi_0\partial_x a_0$
and charge density $\rho=\xi_0\partial_x e_x$.
}
\label{fig:d+is}
\end{figure}

Figure \ref{fig:d+is} shows the data obtained from the self-consistent
solution of the BdG equation for the $d_{xy} + i s$-wave state at zero 
temperature.
Here we present the results for the case where a finite net current is 
running in the system.
Again the order parameter shows strong variation at the surface,
whereby the $d$-wave component is suppressed and the $s$-wave 
component enhanced [Fig.~\ref{fig:d+is}(a)].
Looking at the quasiparticle spectrum in Fig.~\ref{fig:d+is}(b), we
see obvious differences between the chiral and the non-chiral cases.
In both cases there are chiral Andreev bound states below the
ordinary continuous spectra of scattering states.
In the non-chiral case, however, there is no gapless edge mode,
in accordance with the expectation from the index theorem. 
The electromagnetic properties in the Hall bar geometry, shown in
Figs.~\ref{fig:d+is}(c) and (d), are very similar to the chiral case.
Also, the Hall resistance $R_H$ for the parameters indicated in the
figure caption is of similar magnitude.
We thus conclude that for certain surface orientations the measurement
of the transverse voltage in a non-chiral state gives a qualitatively
identical result to a chiral state.
Hence, experiments of this kind on the [110] surface of
high-temperature superconductors, where a low-temperature
time-reversal symmetry breaking phase may be present, would not be
able to give decisive results as to which state is realized, the
non-chiral $d_{x^2-y^2}+is$-wave state or the chiral
$d_{x^2-y^2}+id_{xy}$-wave state.

\section{Conclusions}

We have analyzed in detail the spontaneous Hall effect in time
reversal symmetry breaking quasi-two-dimensional superconductor of
chiral and non-chiral nature.
There are two contributions to the SHE.
One is connected with the Bernoulli (or Lorentz) force due to the
presence of spontaneous surface currents.
The other originates from the presence of an orbital angular momentum
of Cooper pairs.
While the former contribution appears in both types of superconducting
states, there is no angular momentum in the non-chiral case. 
We have shown in our phenomenological treatment that the angular
momentum gives rise to a Chern-Simons-like term in the Lagrangian 
determining the electromagnetism of the superconductor, similar to
derivations based on topological arguments.\cite{Volovik1,Goryo}
Although it was suggested that the Hall response would be, at
least, close to a universal value, our analysis shows that the actual
measurement of the Hall voltage gives a considerably smaller
non-universal value.
Nevertheless, the comparison of the SHE with the ordinary Hall effect
reveals the presence of an intrinsic effective magnetic field
corresponding to the density of Cooper pair magnetic moments.

The effect depends on the presence of spontaneous surface currents.
In our analysis we have restricted ourselves to the case of a perfect,
specularly scattering surface. Rough surfaces with diffuse scattering
would reduce the spontaneous currents and the Hall voltage.
Moreover, domain formation constitutes another obstacle to the
measurement of the SHE, because different domains give contributions
of opposite signs.
Nevertheless, we believe that experimental techniques available
at present are sufficiently accurate to observe the spontaneous Hall
effect.

\acknowledgements

We thank Y.~Maeno, K.~Deguchi, K.~Ishikawa, J.~Goryo, and V.~Yakovenko
for stimulating discussions. 
The work of AF and MS was supported in part by Grant-in-Aid for
Scientific Research on Priority Areas (A) from The Ministry of
Education, Science, Sports and Culture (No.~12046238) and by
Grant-in-Aid for Scientific Research (C) from Japan Society for the
Promotion of Science (No.~10640341).
The work of MM was supported by Grant-in-Aid for Encouragement of
Young Scientists from Japan Society for the Promotion of Science
(No.~10740169) and by Casio Science Promotion Foundation.


\begin{references}

\bibitem{VG} G.\ E.\ Volovik and L.\ P.\ Gor'kov, JETP Lett.\ {\bf
39}, 674 (1984); Sov.\ Phys.\ JETP {\bf 61}, 843 (1985).

\bibitem{Matsu2}  M.\ Matsumoto and H.\ Shiba,
 J.\ Phys.\ Soc.\ Jpn.\ {\bf 64}, 3384 (1995); {\bf 64}, 4867 (1995);
 {\bf 65}, 2194 (1996).

\bibitem{Fogelstrom} M.\ Fogelstr\"om, D.\ Rainer, and J.\ A.\ Sauls,
Phys.\ Rev.\ Lett.\ {\bf 79}, 281 (1997).
 
\bibitem{SigristProgress} M.\ Sigrist, Prog.\ Theor.\ Phys.\ {\bf 99}, 
899 (1998).

\bibitem{HF-REVIEW} R.\ H.\ Heffner and M.\ R.\ Norman, Comments
Condens. Matter Phys.\ {\bf 17}, 361 (1996).

\bibitem{Maeno} Y.\ Maeno, H.\ Hashimoto, K.\ Yoshida, S.\ Nishizaki,
T.\ Fujita, J.\ G.\ Bednorz, and F.\ Lichtenberg, Nature (London)
{\bf 372}, 532 (1994).

\bibitem{RSB} T.\ M.\ Rice and M.\ Sigrist, J.\ Phys.\ Condens.\
Matter {\bf 7}, L643 (1995);
 G.\ Baskaran, Physica B {\bf 223}-{\bf 224}, 490 (1996).

\bibitem{Leggett} A.\ J.\ Leggett, Rev.\ Mod.\ Phys.\ {\bf 47}, 331
(1975). 

\bibitem{Volovik1} G.\ E.\ Volovik, Sov.\ Phys.\ JETP {\bf 67}, 1804
(1988). 

\bibitem{Volovik2} G.\ E.\ Volovik, JETP Lett.\ {\bf 66}, 522 (1997).

\bibitem{Luke} G.\ M.\ Luke, Y.\ Fudamoto, K.\ M.\ Kojima,
M.\ I.\ Larkin, J.\ Merrin, B.\ Nachumi, Y.\ J.\ Uemura, Y.\ Maeno,
Z.\ Q.\ Mao, Y.\ Mori, H.\ Nakamura, and M.\ Sigrist, Nature (London)
{\bf 394}, 558 (1998).

\bibitem{Ishida} K.\ Ishida, H.\ Mukuda, Y.\ Kitaoka, K.\ Asayama,
Z.\ Q.\ Mao, Y.\ Mori, and Y.\ Maeno, Nature (London) {\bf 396}, 658
(1998).

\bibitem{PhysicaC} M.\ Sigrist,
D.\ Agterberg, A.\ Furusaki, C.\ Honerkamp, K.\ K.\ Ng, T.\ M.\ Rice,
and M.\ E.\ Zhitomirsky, Physica C {\bf 317-318}, 134 (1999).

\bibitem{Buch} L.\ J.\ Buchholtz and G.\ Zwicknagle, Phys.\ Rev.\ B
{\bf 23}, 5788 (1981). 

\bibitem{Yamashiro} M.\ Yamashiro, Y.\ Tanaka, and S.\ Kashiwaya,
Phys.\ Rev.\ B {\bf 56}, 7847 (1997).

\bibitem{Honerkamp}
C.\ Honerkamp and M.\ Sigrist, J.\ Low Temp.\ Phys.\ {\bf 111},
895 (1998).

\bibitem{Matsu1} M.\ Matsumoto and M.\ Sigrist,
 J.\ Phys.\ Soc.\ Jpn.\ {\bf 68}, 994 (1999).

\bibitem{LAUBE} F.\ Laube, G.\ Goll, H.\ v.\ L\"ohneysen,
 M.\ Fogelstr\"om, and F.\ Lichtenberg, Phys.\ Rev.\ Lett.\ {\bf 84},
 1595 (2000).

\bibitem{Goryo} J.\ Goryo and K.\ Ishikawa, Phys.\ Lett.\ A {\bf 260}, 
294 (1999).

\bibitem{Senthil} T.\ Senthil, J.\ B.\ Marston, and M.\ P.\ A.\
Fisher, Phys.\ Rev.\ B {\bf 60}, 4245 (1999).

\bibitem{Read} N.\ Read and D.\ Green, Phys.\ Rev.\ B {\bf 61}, 10267
(2000). 

\bibitem{Yakovenko} G.\ E.\ Volovik and V.\ M.\ Yakovenko,
 J.\ Phys.: Condens.\ Matter {\bf 1}, 5263 (1989).

\bibitem{SigristUeda} M.\ Sigrist and K.\ Ueda, Rev.\ Mod.\ Phys.\
{\bf 63}, 239 (1991).

\bibitem{Samokhin} V.\ P.\ Mineev and K.\ V.\ Samokhin,
{\it Introduction to Unconventional Superconductivity}, (Gordon and
Breach, Amsterdam, 1999).

\bibitem{MINEEV} G.\ E.\ Volovik and V.\ P.\ Mineev,
 Sov.\ Phys.\ JETP {\bf 54}, 524 (1981).

\bibitem{CROSS} M.\ C.\ Cross, J.\ Low Temp.\ Phys.\ {\bf 21}, 525
(1975); {\bf 26}, 165 (1977).

\bibitem{MERMIN} N.\ D.\ Mermin and P.\ Muzikar,
 Phys.\ Rev.\ B {\bf 21}, 980 (1980).

\bibitem{Volovik3} G.\ E.\ Volovik, JETP Lett.\ {\bf 61}, 958 (1995).

\bibitem{LONDON} F.\ London, {\it Superfluids}, (John Wiley \& Sons, 
New York, 1950), Vol.~1 Sec.~B. 

\bibitem{STAAS} A.\ G.\ Van Vijfeijken and F.\ A.\ Staas,
 Phys.\ Lett.\ {\bf 12}, 175 (1964). 

\bibitem{RICK} G.\ Rickayzen, Proc.\ Phys.\ Soc.\ (Solid State
Phys.)(GB) {\bf 2}, 1334 (1969).

\bibitem{MORRIS} T.\ D.\ Morris and J.\ B.\ Brown,
 Physica {\bf 55}, 760 (1971).

\bibitem{d+is} This is equivalent to the $d_{x^2-y^2}+is$-wave case
for the [110]-surfaces.


\end{references}
\end{document}